\documentclass[aps,pra,superscriptaddress,twocolumn,a4paper,reprint,noeprint]{revtex4-1}
\usepackage[utf8]{inputenc}
\usepackage{graphicx}
\usepackage{amsmath}
\usepackage{amsfonts}
\usepackage{amssymb}
\usepackage{bbm}
\usepackage[usenames,dvipsnames,svgnames,table]{xcolor}
\usepackage[USenglish]{babel}
\usepackage{hyperref}
\usepackage{grffile}
\usepackage[normalem]{ulem}

\def\sgn{\mathop{\textrm{sgn}}}
\def\Tr{\mathop{\textrm{Tr}}}

\newcommand{\bsigma}{\mbox{\boldmath$\sigma$}}
\newcommand{\bpi}{\mbox{\boldmath$\pi$}}
\newcommand{\bPhi}{\mbox{\boldmath$\Phi$}}
\newcommand{\bnabla}{\mbox{\boldmath$\nabla$}}

\newcommand{\ds}{\displaystyle}
\newcommand{\extrah}{\rule[-2.5ex]{0ex}{6ex}}

\newcommand{\boldone}{\mathbbm{1} }

\allowdisplaybreaks

\begin{document}

\title{Distortional weak-coupling instability of Bogoliubov Fermi surfaces}

\author{Carsten Timm}
\email{carsten.timm@tu-dresden.de}
\affiliation{Institute of Theoretical Physics, Technische Universit\"at Dresden, 01062 Dresden, Germany}
\affiliation{W\"urzburg-Dresden Cluster of Excellence ct.qmat, Technische Universit\"at Dresden, 01062~Dresden, Germany}

\author{P. M. R. Brydon}
\affiliation{Department of Physics and MacDiarmid Institute for Advanced Materials and Nanotechnology, University of Otago, P.O.~Box~56, Dunedin~9054, New Zealand}

\author{Daniel F. Agterberg}
\affiliation{Department of Physics, University of Wisconsin, Milwaukee, WI~53201, USA}

\date{November 10, 2020}

\begin{abstract}
Centrosymmetric multiband superconductors which break time-reversal symmetry generically have two-dimensional nodes, i.e., Fermi surfaces of Bogoliubov quasiparticles. We show that the coupling of the electrons to the lattice always leads to a weak-coupling instability of such a state toward spontaneous breaking of inversion symmetry at low temperatures. This instability is driven by a Cooper logarithm in the internal energy but the order parameter is not superconducting but distortional. We present a comprehensive symmetry analysis and introduce a measure that allows us to compare the strengths of competing distortional instabilities. Moreover, we discuss the instability using an effective single-band model. This framework reveals a duality mapping of the effective model which maps the distortional order parameter onto a superconducting one, providing a natural explanation for the Cooper logarithm and the weak-coupling nature of the instability. Finally, we consider the possibility of a pair-density wave state when inversion symmetry is broken. We find that it can indeed exist but does not affect the instability itself.
\end{abstract}

\maketitle

\section{Introduction}

It has recently been realized that centrosymmetric multiband superconductors breaking time-reversal symmetry generically possess Fermi surfaces of Bogoliubov quasiparticles \cite{ABT17,BAM18}. These Bogoliubov Fermi surfaces (BFSs) are topologically protected by a $\mathbb{Z}_2$ invariant \cite{KST14,ZSW16}, which can be expressed in terms of the Pfaffian of the Bogoliubov--de Gennes Hamiltonian \cite{ABT17,BAM18}. Multiband pairing which breaks time-reversal symmetry has been suggested for various centrosymmetric materials of high current interest, for example iron-based superconductors \cite{GSZ10,NGR12,OCS16,VaC17,ASO17,SBK20,HJW20} and $\mathrm{Sr}_2\mathrm{RuO}_4$~\cite{TaK12,SMB20,CLK20}. A particularly interesting candidate is $\mathrm{U}_{1-x}\mathrm{Th}_x\mathrm{Be}_{13}$ \cite{Ste19}, which shows two transitions in a certain doping range \cite{ORF85,RBS87}, where time-reversal symmetry is broken in the low-temperature phase \cite{HSW90}. Moreover, the observation that the specific heat and the thermal conductivity are linear in temperature under pressure \cite{ZDS04} hint at the existence of BFSs \cite{LBT20}. These results suggest unconventional superconductivity but there is no consensus as to the pairing state in the two phases~\cite{SiR89,Mac18,Ste19}.

The existence of the $\mathbb{Z}_2$ invariant relies on the product of charge conjugation and inversion symmetries ($\mathcal{C}P$) and it is thus natural to ask how stable this state and thus the BFSs are. In particular, a real crystal might deform in such a way that inversion symmetry and thus $\mathcal{C}P$ are broken, which presumably would gap out the BFSs and push spectral weight away from the Fermi energy. This process should reduce the electronic internal energy at the price of increasing the elastic energy. The question arises whether spontaneous breaking of $\mathcal{C}P$ symmetry by distortion is possible and, if so, under what conditions. Recently, Oh and Moon \cite{OhM20} as well as Tamura \textit{et al.}\ \cite{TIH20} have shown that there is a weak-coupling instability for an attractive quasiparticle interaction in an inversion-symmetry-breaking channel, the origin of which is not discussed. The authors' models do not contain lattice distortions, which are the main players here. Very recently, Link and Herbut \cite{LiH20} have investigated the complementary case of time-reversal-symmetry-breaking superconductivity emerging in materials that already break inversion symmetry in the normal state. Their perturbative analysis suggests that such systems often have BFSs for energetic instead of topological reasons.

In this paper, we show that the BFSs in centrosymmetric multiband superconductors are \emph{generically} unstable toward distortions that break inversion symmetry. To be precise, if such a system possesses odd-parity zone-center phonon modes, then there is a distortional weak-coupling instability at zero temperature. This instability results from a Cooper logarithm in the internal energy, i.e., a term proportional to $\Phi^2 \ln \Phi$, where $\Phi$ is the suitably defined magnitude of the distortional order parameter. Notably, the Cooper logarithm here appears for a distortional instability, whereas it is usually characteristic for superconducting instabilities. Its origin in both cases is that a gap opens at the Fermi energy. In the present case, this effect is due to the loss of the $\mathbb{Z}_2$ topological invariant, which protects the BFSs in centrosymmetric systems.

The breaking of inversion symmetry also leads to odd-in-momentum shifts of the quasiparticle bands. We find that remaining symmetries can prevent the gapping or the shifts for certain distortions and certain momenta. If the band shift is larger than the distortional gap the distorted system still possesses BFSs even though there is no $\mathbb{Z}_2$ invariant. It has been known before that superconductors that break both time-reversal and inversion symmetry can have BFSs~\cite{Vol89,TSA17,YuF18,Vol18,LBH20,LiH20}.

In addition, we find that the odd-in-momentum band shifts are irrelevant for the instability to leading order since they do not affect the Cooper logarithm. We derive a relatively simple measure involving the distortional gap in the quasiparticle bands and the normal-state Fermi velocity that allows to assess and compare the strengths of competing distortional instabilities.

In this paper, we make general statements and illustrate them for a particular model system, which is introduced in Sec.\ \ref{sec.model}. In the following sections, a comprehensive symmetry analysis of the distortional instability is presented, based on the magnetic point group, which is advantageous in that it treats generalized rotations and time reversal on equal footing and reveals symmetries of the superconducting state that could otherwise be overlooked. Specifically, uniform distortions are discussed in Sec.\ \ref{sec.distort} and their coupling to the electronic degrees of freedom is analyzed in Sec.\ \ref{sec.coupling}. We then consider the effects of the distortion on the quasiparticle dispersion in Sec.\ \ref{sec.qpdisp} and the instability of the BFSs in Sec.\ \ref{sec.BFS_instab}. Here, we also introduce a method to assess competing distortional instabilities and illustrate it with numerical results.

In Sec.\ \ref{sec.pseudo}, distortions are analyzed in the pseudospin (single-band) picture \cite{ABT17,BAM18}. The pseudospin picture simplifies the symmetry analysis and also permits a duality mapping under which the distortional instability becomes a superconducting instability of the dual model. Finally, the breaking of both time-reversal and inversion symmetries begs the question whether the homogeneous superconducting state is unstable toward a modulated superconducting state, i.e., a pair-density wave (PDW) \cite{SSY17}. In Sec.\ \ref{sec.PDW}, we employ Ginzburg-Landau theory to show that a PDW instability can indeed emerge for some but not all distortion modes. If such an instability is present the wavelength of the PDW diverges at the distortional instability so that the PDW is irrelevant for the weak-coupling instability. Finally, the work is summarized and conclusions are drawn in Sec.~\ref{sec.conclusions}.

\section{Model system}
\label{sec.model}

For illustration, we use the caesium chloride structure, which arguably is the simplest one to exhibit the relevant physics and provides a good approximation for the structure of (doped) $\mathrm{UBe}_{13}$. In $\mathrm{UBe}_{13}$, the uranium atoms and one in 13 beryllium atoms ($\mathrm{Be}2$) form a caesium chloride structure, while the remaining beryllium atoms ($\mathrm{Be}1$) sit on the faces of the cubes formed by $\mathrm{Be}2$, forming cubic boxes around the uranium \cite{BaR49}. Both the caesium chloride structure and $\mathrm{UBe}_{13}$ have the crystallographic point group $O_h$. The magnetic point group of the normal state is $O_h \otimes \{1,\mathcal{T}\}$, where $\mathcal{T}$ denotes time reversal. Since $\{1,\mathcal{T}\}$ appears as a factor group, the magnetic point group is ``gray'' \cite{Lit13,Kop15}. The Hermann--Mauguin notation is $m\bar{3}m1'$, where $1'$ stands for time reversal combined with the spatial identity operation \cite{Lit13}. The group has two irreps $\Gamma_+$ and $\Gamma_-$ for any irrep $\Gamma$ of $O_h$, where the subscript $\pm$ indicates the sign under time reversal.

To be specific, we consider electrons close to a $\Gamma_8$ band-touching point described by an effective spin of length $J=3/2$ \cite{BWW16,ABT17,TSA17,YXW17,SRV17,BAM18,OhM20,TIH20}. The Hilbert space of the local degrees of freedom is four dimensional. A basis of Hermitian matrices spanning this space can be constructed from the standard spin-$3/2$ matrices. These basis matrices are irreducible tensor operators of irreps of the magnetic point group and are listed in Table \ref{tab.basis}, together with their irreps. The normal-state Hamiltonian can be written as
\begin{equation}
H_N(\mathbf{k}) = \sum_{n=0}^5 c_n(\mathbf{k})\, h_n .
\label{eq.HN.3}
\end{equation}
The momentum-dependent functions $c_n(\mathbf{k})$ must transform like
the corresponding matrices $h_n$ so that the full Hamiltonian is
invariant under all symmetry transformations (i.e., belongs to the trivial irrep $A_{1g+}$). The specific functions $c_n(\mathbf{k})$ are given in Sec.\ \ref{sub.numerics}. Only the specific six basis matrices belonging to $g+$ irreps can appear because $g-$ irreps do not have momentum basis functions. Given the normal-state Hamiltonian $H_N(\mathbf{k})$, the Bogoliubov--de Gennes (BdG) Hamiltonian for the superconducting state reads as
\begin{equation}
\mathcal{H}(\mathbf{k}) = \begin{pmatrix}
    H_N(\mathbf{k}) & \Delta(\mathbf{k}) \\
    \Delta^\dagger(\mathbf{k}) & -H_N^T(-\mathbf{k})
  \end{pmatrix} , 
\label{eq.HBdG.3}
\end{equation}
where $\bullet^T$ denotes the transpose.

\begin{table}
\caption{\label{tab.basis}Basis matrices of the four-dimensional Hilbert space of an effective spin of length $J=3/2$. The matrices have been normalized in such a way that $\Tr AB=4\delta_{AB}$. The last two columns show the corresponding irreps for the magnetic point groups $m\bar{3}m1'$ and $4/mm'm'$, respectively. The matrices belonging to $g+$ irreps of $m\bar{3}m1'$ can appear in the Hamiltonian for the normal state without distortion and are denoted by $h_0$, \dots, $h_5$. The table does not specify the order and sign of the components of the $E_g$ doublets of $4/mm'm'$. Using $(xz,yz)$ as the template, the $E_g$ doublets are, suppressing common factors, $(J_x, J_y)$, $(J_zJ_x + J_xJ_z, J_yJ_z + J_zJ_y)$, $(J_x^3, J_y^3)$, and $(J_x(J_y^2-J_z^2)+(J_y^2-J_z^2)J_x, - J_y(J_z^2-J_x^2)-(J_z^2-J_x^2)J_y)$.}
\begin{ruledtabular}
\begin{tabular}{ccc}
 & \multicolumn{2}{c}{irrep} \\[-0.7ex]
\raisebox{1.8ex}[-1.8ex]{basis matrix} & \rule[-1ex]{0ex}{4ex} $m\bar{3}m1'$ & $4/mm'm'$ \\ \hline\hline
$\boldone \equiv h_0$ & $A_{1g+}$ & $A_{1g}$ \\ \hline
\extrah $\ds\frac{2}{\sqrt{5}}\, J_x$ & & \\
\extrah $\ds\frac{2}{\sqrt{5}}\, J_y$ & $T_{1g-}$ & \raisebox{2.5ex}[-2.5ex]{$E_g$} \\ \cline{3-3}
\extrah $\ds\frac{2}{\sqrt{5}}\, J_z$ & & $A_{1g}$ \\ \hline
\extrah $\ds\frac{1}{\sqrt{3}}\, (J_yJ_z + J_zJ_y) \equiv h_1$ & & \\
\extrah $\ds\frac{1}{\sqrt{3}}\, (J_zJ_x + J_xJ_z) \equiv h_2$ & $T_{2g+}$ & \raisebox{2.5ex}[-2.5ex]{$E_g$} \\ \cline{3-3}
\extrah $\ds\frac{1}{\sqrt{3}}\, (J_xJ_y + J_yJ_x) \equiv h_3$ & & $B_{2g}$ \\ \hline
\extrah $\ds\frac{1}{\sqrt{3}}\, (J_x^2 - J_y^2) \equiv h_4$ & & $B_{1g}$ \\ \cline{3-3}
\extrah $\ds\frac{1}{3}\, (2J_z^2 - J_x^2 - J_y^2) \equiv h_5$ & \raisebox{2.5ex}[-2.5ex]{$E_{g+}$} & $A_{1g}$ \\ \hline
\extrah $\ds\frac{2}{\sqrt{3}}\, (J_xJ_yJ_z + J_zJ_yJ_x)$ & $A_{2g-}$ & $B_{2g}$ \\ \hline 
\extrah $\ds\frac{8}{\sqrt{365}}\, J_x^3$ & & \\
\extrah $\ds\frac{8}{\sqrt{365}}\, J_y^3$ & $T_{1g-}$ & \raisebox{2.5ex}[-2.5ex]{$E_g$} \\ \cline{3-3}
\extrah $\ds\frac{8}{\sqrt{365}}\, J_z^3$ & & $A_{1g}$ \\ \hline
\extrah $\ds\frac{1}{\sqrt{3}}\, \big[ J_x(J_y^2-J_z^2) + (J_y^2-J_z^2)J_x \big]$ & & \\
\extrah $\ds\frac{1}{\sqrt{3}}\, \big[ J_y(J_z^2-J_x^2) + (J_z^2-J_x^2)J_y \big]$ & $T_{2g-}$ & \raisebox{2.5ex}[-2.5ex]{$E_g$} \\ \cline{3-3}
\extrah $\ds\frac{1}{\sqrt{3}}\, \big[ J_z(J_x^2-J_y^2) + (J_x^2-J_y^2)J_z \big]$ & & $B_{1g}$
\end{tabular}
\end{ruledtabular}
\end{table}

\begin{figure}[t!]
\includegraphics[width=0.8\columnwidth]{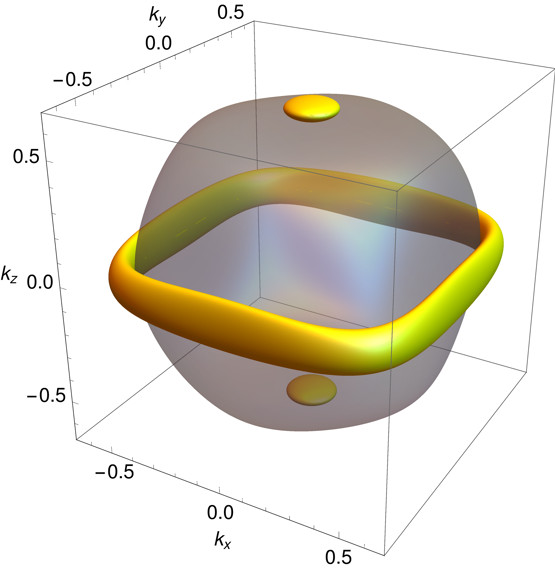}
\caption{\label{fig.BFS}Normal-state Fermi surface (semitransparent gray) and BFSs (yellow) for the specific model with superconducting chiral $T_{2g}$ order parameter $\Delta\,(1,i,0)$. The parameters of the model are specified in Sec.~\ref{sub.numerics}.}
\end{figure}

We are interested in superconducting states that break at least time-reversal symmetry and may also break spatial point-group symmetries. We specifically consider the chiral $T_{2g}$ pairing state with the order parameter $\Delta\,(1,i,0)$ studied in Refs.\ \cite{ABT17,BAM18}. This state has \textit{s}-wave momentum dependence (local pairing) but nontrivial orbital structure. Other pairing states are briefly commented upon in Sec.\ \ref{sec.conclusions}. The BFS of this model is depicted in Fig.\ \ref{fig.BFS}. Explicit evaluation of the transformations of the BdG Hamiltonian shows that the magnetic point group is reduced to $D_{4h}(C_{4h}) = 4/mm'm' = \text{\#15.6.58}$ \cite{Lit13}, where the notation $D_{4h}(C_{4h})$ means that the group contains the elements of $C_{4h}$ and in addition the elements of $D_{4h}\setminus C_{4h}$ multiplied by time reversal. Specifically, the twofold rotations with rotation vector in the $xy$ plane and the mirror reflections with normal vector in the $xy$ plane are multiplied by time reversal. The irreps are formally the same as for $D_{4h}$ since the groups are isomorphic~\cite{ErH20}.

Symmetries of the BdG Hamiltonian can be treated in essentially the same way as symmetries of the normal-state Hamiltonian, except that the phase factor of unitary matrices, which is arbitrary for the normal state, has to be fixed. This reflects that superconductivity breaks a global $\mathrm{U}(1)$ invariance down to $\mathbb{Z}_2$. A general discussion is given in Appendix~\ref{app.symmetry}.

It will prove useful to classify the basis matrices according to irreps of the magnetic point group of the superconducting state. The results are also included in Table \ref{tab.basis}. Care must be taken for odd powers of spin since some of the group elements involve time reversal. For example, $J_z$ is odd under twofold rotation about the \textit{x} axis and belongs to the irrep $A_{2g}$ of $D_{4h}$. However, $D_{4h}(C_{4h}) = 4/mm'm'$ contains this rotation multiplied by time reversal. $J_z$ is even under this combined transformation and belongs to the irrep $A_{1g}$ of $D_{4h}(C_{4h})$.

\section{Distortion modes}
\label{sec.distort}

The first step of the analysis is the identification of all uniform distortion modes that break inversion symmetry. These are the modes that transform according to odd-parity ($u$) irreps of the point group. The uniform distortion modes are the optical phonons at $\mathbf{k}\to 0$. The result is a list of all inversion-symmetry-breaking deformation modes with their associated irreps.

Not all structures with a nontrivial basis have zone-center phonons belonging to $u$ irreps. For example, the diamond structure has a two-atom basis and thus three optical-phonon branches. These transform according to the irrep $T_{2g}$ of the point group $O_h$ \cite{Bir63} and the irrep $T_{2g+}$ of the magnetic point group. Since $T_{2g+}$ is three dimensional there are no other uniform distortion modes, in particular no odd-parity ones. This means that the inversion symmetry of the diamond structure cannot be broken by a uniform distortion. Of course, nonuniform distortions, which also break translation symmetry, are possible. If they are commensurate they are easily described in the same framework by considering a superlattice in which the distortion is uniform. We return to this point in Sec.~\ref{sec.conclusions}.

The caesium chloride structure also has a two-atom basis and thus three optical-phonon branches. However, the zone-center optical phonons here transform according to $T_{1u+}$ \cite{GBK65}, i.e., like a spatial vector, which is plausible since the distortions generate an electric dipole moment of each unit cell. We denote the displacement vector by $\mathbf{p}=(p_x,p_y,p_z)$, which serves as the distortional order parameter. The caesium chloride structure thus has three degenerate distortion modes that break inversion symmetry \cite{endnote.polar}. $\mathrm{UBe}_{13}$ inherits these modes since the uranium and $\mathrm{Be}2$ sublattices form a caesium chloride structure but of course has many additional ones.

So far, we have not considered the symmetry reduction due to superconductivity. Using a table of basis functions for $D_{4h}$ and the fact that the displacement vector is even under time reversal, we find that the three degenerate distortion modes of the caesium chloride are split into a doublet $(p_x,p_y)$, which transforms according to $E_u$, and a singlet $p_z$, which belongs to $A_{2u}$.

\section{Coupling to the electronic system}
\label{sec.coupling}

The second step is to set up the symmetry-allowed terms $H_\text{dist}(\mathbf{k})$ in the electronic Hamiltonian for every odd-parity distortion mode. Since we are looking for a weak-coupling instability we may restrict ourselves to terms linear in the distortions. We expand $H_\text{dist}(\mathbf{k})$ in the basis matrices in Table \ref{tab.basis}. Standard methods of group theory are used to find the allowed terms: the combination of distortional order parameters, basis matrices, and momentum-dependent structure factors must be invariant under the magnetic point group, i.e., belong to the trivial irrep. Note that we must include the distortional order parameter in the transformations to find its coupling to the electronic degrees of freedom. If the distortion is kept fixed it breaks lattice symmetries and thereby reduces the point group.

For each distortion mode or multiplet of modes belonging to some irrep $\Gamma_\text{dist}$ and each basis matrix or multiplet of matrices belonging to some irrep $\Gamma_h$, we are searching for possible momentum basis functions so that their combination has the full symmetry of the point group. This is most easily done by reducing the product representation $\Gamma_\text{dist} \otimes \Gamma_h$. Suitable momentum basis functions must belong to one of the irreps appearing in the reduction. In our example, we have distortions belonging to $T_{1u+}$ of $m\bar{3}m1'$ in the normal state and to $E_u$ and $A_{2u}$ of $4/mm'm'$ in the superconducting state. All basis matrices belong to $g$ irreps so that the product representation is parity odd. Since momentum is also odd under parity the structure factor must be odd in momentum, i.e., only momentum basis functions belonging to $u$ irreps can occur. Since momentum is also odd under time reversal only $u-$ irreps of $m\bar{3}m1'$ occur. In more detail, we obtain the reductions given in Table \ref{tab.reduce1} for the normal state and in Table \ref{tab.reduce2} for the superconducting state.

\begin{table}
\caption{\label{tab.reduce1}Reduction of product representations relevant for the electronic Hamiltonian $H_\text{dist}(\mathbf{k})$ due to distortions of a caesium-chloride system in the normal state. $u+$ irreps do not have momentum basis functions and thus do not lead to allowed terms in $H_\text{dist}(\mathbf{k})$. These irreps are set in parentheses.}
\begin{ruledtabular}
\begin{tabular}{ccl}
$\Gamma_\text{dist}$ & $\Gamma_h$ & product representation \\ \hline\hline
 & $A_{1g+}$ & $(T_{1u+})$ \\
 & $A_{2g-}$ & $T_{2u-}$ \\
 & $E_{g+}$ & $(T_{1u+} \oplus T_{2u+})$ \\
\raisebox{1.2ex}[-1.2ex]{$T_{1u+}$} & $T_{1g-}$ & $A_{1u-} \oplus E_{u-} \oplus T_{1u-} \oplus T_{2u-}$ \\
 & $T_{2g+}$ & $(A_{2u+} \oplus E_{u+} \oplus T_{1u+} \oplus T_{2u+})$ \\
 & $T_{2g-}$ & $A_{2u-} \oplus E_{u-} \oplus T_{1u-} \oplus T_{2u-}$
\end{tabular}
\end{ruledtabular}
\end{table}

\begin{table}
\caption{\label{tab.reduce2}Reduction of product representations relevant for the electronic Hamiltonian $H_\text{dist}(\mathbf{k})$ due to distortions in the state with $T_{2g}$, $\Delta(1,i,0)$} superconductivity.
\begin{ruledtabular}
\begin{tabular}{ccl}
$\Gamma_\text{dist}$ & $\Gamma_h$ & product representation \\ \hline\hline
 & $A_{1g}$ & $E_u$ \\
 & $B_{1g}$ & $E_u$ \\
\raisebox{1.2ex}[-1.2ex]{$E_u$} & $B_{2g}$ & $E_u$ \\
 & $E_g$ & $A_{1u} \oplus A_{2u} \oplus B_{1u} \oplus B_{2u}$ \\ \hline
 & $A_{1g}$ & $A_{2u}$ \\
 & $B_{1g}$ & $B_{2u}$ \\
\raisebox{1.2ex}[-1.2ex]{$A_{2u}$} & $B_{2g}$ & $B_{1u}$ \\
 & $E_g$ & $E_u$
\end{tabular}
\end{ruledtabular}
\end{table}

It is useful to obtain the lowest-order polynomial momentum basis functions for the $u$ irreps of $4/mm'm'$. These are distinct from the basis functions for $D_{4h} = 4/mmm$ since momentum transforms differently under the transformations involving time reversal. For the two-dimensional irrep $E_u$, we take $(x,y)$ as the template defining the first and second components. The resulting basis functions are shown in Table \ref{tab.kubasis}.

\begin{table}
\caption{\label{tab.kubasis}Lowest-order momentum basis functions of $u$ irreps of the magnetic point group $4/mm'm'$.}
\begin{ruledtabular}
\begin{tabular}{cl}
$\Gamma$ & basis functions \\ \hline\hline
$A_{1u}$ & $k_z$ \\
$A_{2u}$ & $k_xk_yk_z\, (k_x^2-k_y^2)$ \\
$B_{1u}$ & $k_z\, (k_x^2-k_y^2)$ \\
$B_{2u}$ & $k_xk_yk_z$ \\
$E_u$ & $(k_y,-k_x)$
\end{tabular}
\end{ruledtabular}
\end{table}

Tables \ref{tab.basis}--\ref{tab.reduce2} show that some basis matrices cannot occur in $H_\text{dist}(\mathbf{k})$ in the normal state but become allowed in the superconducting state. The reason for this is not the presence of superconductivity but the breaking of time-reversal symmetry. As an example, consider the term $(k_xp_y - k_yp_x)\,\boldone$. It is allowed in the superconducting state since both $(p_x,p_y)$ and $(k_y,-k_x)$ are $E_u$ doublets. However, it cannot appear in the normal state since it is odd under time reversal. Since such terms are induced by the time-reversal-symmetry-breaking superconductivity their prefactors must vanish if the pairing amplitude $\Delta$ vanishes. Invariance under global phase rotations of $\Delta$ as well as perturbative arguments imply that the prefactors will be proportional to $|\Delta|^2$ to leading order \cite{ABT17,BAM18}. If the superconducting energy scale $|\Delta|$ is small compared to normal-state energy scales, then distortional terms of this type are expected to be small compared to terms that are also allowed in the normal state.

\begin{table}
\caption{\label{tab.EuEg.products}Combinations of $(v_1,v_2)$ transforming according to $E_u$ and $(w_1,w_2)$ transforming according to $E_g$ that belong to the four one-dimensional $u$ irreps of $4/mm'm'$. The third column shows the corresponding expressions for $(k_y,-k_x)$ and $(J_x,J_y)$, while the fourth shows them for $(p_x,p_y)$ and $(J_x,J_y)$.}
\begin{ruledtabular}
\begin{tabular}{cccc}
$A_{1u}$ & $v_1w_2 - v_2w_1$ & $k_y J_y + k_x J_x$ & $p_x J_y - p_y J_x$ \\
$A_{2u}$ & $v_1w_1 + v_2w_2$ & $k_y J_x - k_x J_y$ & $p_x J_x + p_y J_y$ \\
$B_{1u}$ & $v_1w_2 + v_2w_1$ & $k_y J_y - k_x J_x$ & $p_x J_y + p_y J_x$ \\
$B_{2u}$ & $v_1w_1 - v_2w_2$ & $k_y J_x + k_x J_y$ & $p_x J_x - p_y J_y$
\end{tabular}
\end{ruledtabular}
\end{table}

\begin{table}
\caption{\label{tab.EuEu.products}Combinations of $(v_1,v_2)$ and $(w_1,w_2)$ transforming according to $E_u$ that belong to the four one-dimensional $g$ irreps of $4/mm'm'$. The last column shows the corresponding expressions for $(k_y,-k_x)$ and $(p_x,p_y)$.}
\begin{ruledtabular}
\begin{tabular}{ccc}
$A_{1g}$ & $v_1w_1 + v_2w_2$ & $k_y p_x - k_x p_y$ \\
$A_{2g}$ & $v_1w_2 - v_2w_1$ & $k_x p_x + k_y p_y$ \\
$B_{1g}$ & $v_1w_1 - v_2w_2$ & $k_y p_x + k_x p_y$ \\
$B_{2g}$ & $v_1w_2 + v_2w_1$ & $-k_x p_x + k_y p_y$
\end{tabular}
\end{ruledtabular}
\end{table}

For illustration, we construct the complete distortion terms in the Hamiltonian using Tables \ref{tab.basis}--\ref{tab.reduce2}. This is straightforward for one-dimensional irreps but care is needed for the products of the two-dimensional irreps $E_u$ and $E_g$. These occur because $(p_x,p_y)$ as well as $(k_y,-k_x)$ are $E_u$ doublets, while certain pairs of basis matrices, for example $(J_x,J_y)$, are $E_g$ doublets. We take $(x,y)$ and $(xz,yz)$ as templates for the ordering of basis functions of $E_u$ and $E_g$, respectively. The correct combinations belonging to the four one-dimensional irreps in $E_u \otimes E_g = A_{1u} \oplus A_{2u} \oplus B_{1u} \oplus B_{2u}$ can be inferred form these templates. They are given in Table \ref{tab.EuEg.products}. The corresponding combinations belonging to the one-dimensional irreps in $E_u \otimes E_u = A_{1g} \oplus A_{2g} \oplus B_{1g} \oplus B_{2g}$ are given in Table \ref{tab.EuEu.products}. The resulting contributions to the Hamiltonian for distortions $p_z$ read as
\begin{widetext}
\begin{align}
H_{p_z}(\mathbf{k}) &= \bigg[ d_{A_{2u}}^1(\mathbf{k})\, J_z
  + d_{A_{2u}}^2(\mathbf{k})\, J_z^3
  + d_{B_{2u}}^1(\mathbf{k})\, [J_z(J_x^2-J_y^2) + (J_x^2-J_y^2)J_z] \nonumber \\
& \quad{}+ d_{B_{1u}}^1(\mathbf{k})\, (J_xJ_yJ_z + J_zJ_yJ_x)
  + \vec d_{E_u}^{\:1}(\mathbf{k}) \cdot \begin{pmatrix} J_x \\ J_y \end{pmatrix}
  + \vec d_{E_u}^{\:2}(\mathbf{k}) \cdot \begin{pmatrix} J_x^3 \\[0.5ex] J_y^3 \end{pmatrix} \nonumber \\
& \quad{}+ \vec d_{E_u}^{\:3}(\mathbf{k}) \cdot \begin{pmatrix} J_x(J_y^2-J_z^2) + (J_y^2-J_z^2)J_x \\[0.5ex]
  - J_y(J_z^2-J_x^2) - (J_z^2-J_x^2)J_y \end{pmatrix} \nonumber \\
& \quad{}+ |\Delta|^2\, e_{A_{2u}}^1(\mathbf{k})\, \boldone 
  + |\Delta|^2\, e_{A_{2u}}^2(\mathbf{k})\, (2J_z^2-J_x^2-J_y^2)
  + |\Delta|^2\, e_{B_{2u}}^1(\mathbf{k})\, (J_x^2-J_y^2) \nonumber \\
& \quad{}+ |\Delta|^2\, e_{B_{1u}}^1(\mathbf{k})\, (J_xJ_y+J_yJ_x)
  + |\Delta|^2\, \vec e_{E_u}^{\:1}(\mathbf{k}) \cdot \begin{pmatrix} J_zJ_x + J_xJ_z \\
    J_yJ_z + J_zJ_y \end{pmatrix} \bigg]\, p_z ,
\label{eq.Hdist.pz}
\end{align}
while for distortions $(p_x,p_y)$ we obtain
\begin{align}
H_{p_x,p_y}(\mathbf{k}) &= \bigg[ \vec d_{E_u}^{\:4}(\mathbf{k})\, J_z
  + \vec d_{E_u}^{\:5}(\mathbf{k})\, J_z^3
  + \vec d_{E_u}^{\:6}(\mathbf{k}) \begin{pmatrix} 1 & 0 \\ 0 & -1 \end{pmatrix}
    [J_z(J_x^2-J_y^2) + (J_x^2-J_y^2)J_z]
  \nonumber \\
& \quad{}+ \vec d_{E_u}^{\:7}(\mathbf{k})
    \begin{pmatrix} 0 & 1 \\ 1 & 0 \end{pmatrix} (J_xJ_yJ_z + J_zJ_yJ_x)
  + d_{A_{1u}}^1(\mathbf{k})\, (J_y, -J_x)
  + d_{A_{1u}}^2(\mathbf{k})\, (J_y^3, -J_x^3) \nonumber \\
& \quad{}+ d_{A_{1u}}^3(\mathbf{k})\,
    (-J_y(J_z^2-J_x^2)-(J_z^2-J_x^2)J_y, -J_x(J_y^2-J_z^2)-(J_y^2-J_z^2)J_x)
  + d_{A_{2u}}^3(\mathbf{k})\, (J_x, J_y) \nonumber \\
& \quad{}+ d_{A_{2u}}^4(\mathbf{k})\, (J_x^3, J_y^3)
  + d_{A_{2u}}^5(\mathbf{k})\, (J_x(J_y^2-J_z^2)+(J_y^2-J_z^2)J_x,
    -J_y(J_z^2-J_x^2)-(J_z^2-J_x^2)J_y)
  \nonumber \\
& \quad{}+ d_{B_{1u}}^2(\mathbf{k})\, (J_y, J_x)
  + d_{B_{1u}}^3(\mathbf{k})\, (J_y^3, J_x^3) \nonumber \\
& \quad{}+ d_{B_{1u}}^4(\mathbf{k})\, (-J_y(J_z^2-J_x^2)-(J_z^2-J_x^2)J_y,
    J_x(J_y^2-J_z^2)+(J_y^2-J_z^2)J_x)
  + d_{B_{2u}}^2(\mathbf{k})\, (J_x, -J_y) \nonumber \\
& \quad{}+ d_{B_{2u}}^3(\mathbf{k})\, (J_x^3, -J_y^3)
  + d_{B_{2u}}^4(\mathbf{k})\, (J_x(J_y^2-J_z^2)+(J_y^2-J_z^2)J_x,
    J_y(J_z^2-J_x^2)+(J_z^2-J_x^2)J_y)
  \nonumber \\
& \quad{}+ |\Delta|^2\, \vec e_{E_u}^{\:2}(\mathbf{k})\, \boldone 
  + |\Delta|^2\, \vec e_{E_u}^{\:3}(\mathbf{k})\, (2J_z^2-J_x^2-J_y^2)
  + |\Delta|^2\, \vec e_{E_u}^{\:4}(\mathbf{k})
    \begin{pmatrix} 1 & 0 \\ 0 & -1 \end{pmatrix} (J_x^2-J_y^2)
  \nonumber \\
& \quad{}+ |\Delta|^2\, \vec e_{E_u}^{\:5}(\mathbf{k})
    \begin{pmatrix} 0 & 1 \\ 1 & 0 \end{pmatrix} (J_xJ_y+J_yJ_x)
  + |\Delta|^2\, e_{A_{1u}}^1(\mathbf{k})\, (J_yJ_z+J_zJ_y, -J_zJ_x-J_xJ_z)
  \nonumber \\
& \quad{}+ |\Delta|^2\, e_{A_{2u}}^3(\mathbf{k})\, (J_zJ_x+J_xJ_z, J_yJ_z+J_zJ_y)
  + |\Delta|^2\, e_{B_{1u}}^2(\mathbf{k})\, (J_yJ_z+J_zJ_y, J_zJ_x+J_xJ_z)
  \nonumber \\
& \quad{}+ |\Delta|^2\, e_{B_{2u}}^2(\mathbf{k})\, (J_zJ_x+J_xJ_z, -J_yJ_z-J_zJ_y)
  \bigg] \cdot \begin{pmatrix} p_x \\ p_y \end{pmatrix} ,
\label{eq.Hdist.pxpy}
\end{align}
\end{widetext}
where the functions $d_\Gamma^l(\mathbf{k})$, $\vec d_{E_u}^{\:l}(\mathbf{k})$, $e_\Gamma^l(\mathbf{k})$, and $\vec e_{E_u}^{\:l}(\mathbf{k})$ transform as momentum basis functions of the indicated irreps of $4/mm'm'$. Different functions belonging to the same irrep are enumerated by the superscript $l$, and the notation $d$ and $e$ distinguishes between terms that are present in both the normal and superconducting and only in the superconducting state, respectively. For terms that only exist in the superconducting state, the leading dependence on the pairing amplitude $\Delta$ is made explicit. Vector arrows, two-component row and column vectors, $2\times 2$ matrices, and the dot product refer to two-dimensional irreps.

In this way, all symmetry-allowed terms in $H_\text{dist}(\mathbf{k})$ associated with all odd-parity distortion modes can be constructed for any system. The corresponding terms in the BdG Hamiltonian are then obtained as
\begin{equation}
\mathcal{H}_\text{dist}(\mathbf{k}) = \begin{pmatrix}
    H_\text{dist}(\mathbf{k}) & 0 \\
    0 & -H_\text{dist}^T(-\mathbf{k})
  \end{pmatrix} .
\end{equation}

\begin{figure*}
\raisebox{1ex}[-1ex]{(a)}\includegraphics[scale=0.385]{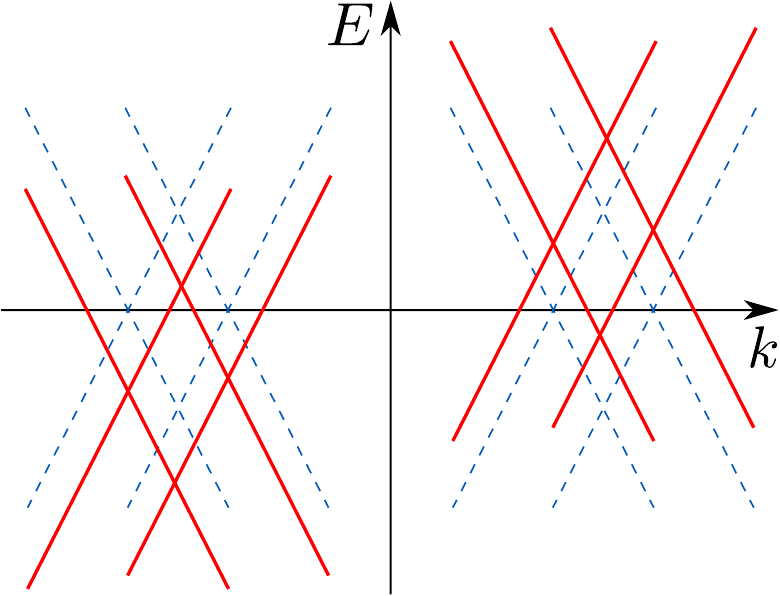}\qquad%
\raisebox{1ex}[-1ex]{(b)}\includegraphics[scale=0.385]{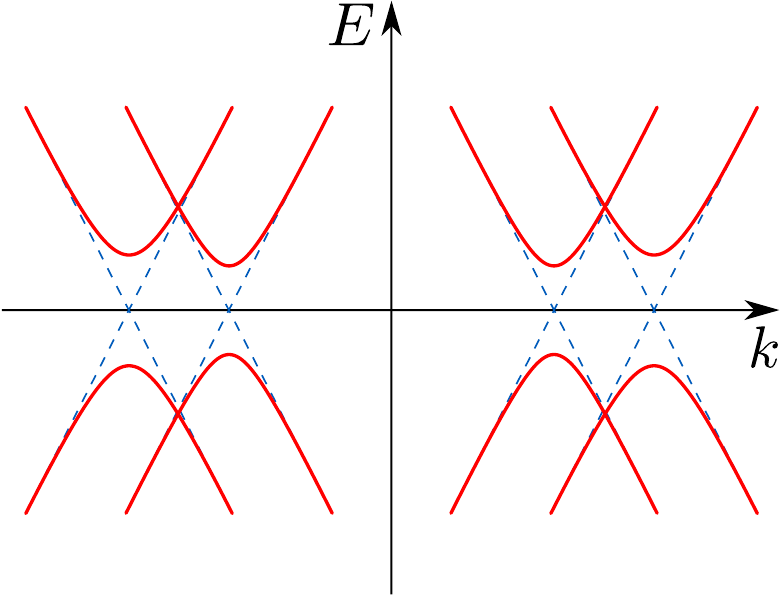}\qquad%
\raisebox{1ex}[-1ex]{(c)}\includegraphics[scale=0.385]{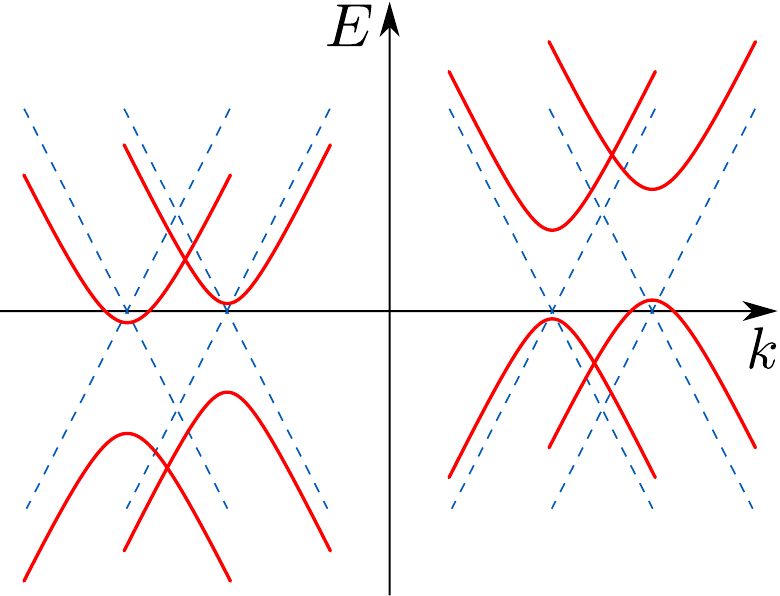}
\caption{\label{fig.sketch}Sketches of the consequences of distortions: (a) shifts of quasiparticle bands, (b) splitting of band crossings, and (c) combination of both. The light blue dashed lines denote the quasiparticle bands of the undistorted superconductor. Their crossings with $E=0$ form the BFSs of the centrosymmetric superconductor. Note that the distorted system retains BFSs when the distortional gap is smaller than the band shift.}
\end{figure*}

\section{Effects on the quasiparticle dispersion}
\label{sec.qpdisp}

The distortional terms $\mathcal{H}_\mathrm{dist}(\mathbf{k})$ in the BdG Hamiltonian break inversion symmetry if the distortion is held fixed and are therefore expected to have two distinct effects: odd-in-momentum shifts of the quasiparticle bands and splitting of band crossings. They correspond to the \emph{Pomeranchuk} and \emph{Cooper} instabilities, respectively, of BFSs discussed by Tamura \textit{et al.}\ \cite{TIH20} for a purely electronic model. The cases of pure shifts and pure splittings as well as their combination are sketched in Fig.\ \ref{fig.sketch}. We now discuss these effects in some detail.

On the one hand, the distorted superconducting system has neither
inversion symmetry $P$ nor time-reversal symmetry $\mathcal{T}$. The
only surviving global symmetry is charge conjugation $\mathcal{C}$,
which guarantees that for any quasiparticle state of energy $E$ at
momentum $\mathbf{k}$, there is another quasiparticle state of energy
$-E$ at $-\mathbf{k}$. However, there is no symmetry that requires the
presence of a state of energy $E$ at $-\mathbf{k}$. This means that
the quasiparticle band structure is generally not even in
momentum. Since it approaches an even limit for vanishing distortions,
there must be corrections that are odd in $\mathbf{k}$. In fact, for
our model, all occurring momentum basis functions are odd so that
$\mathcal{H}_\mathrm{dist}(\mathbf{k})$ is odd. However, this is only
true because the local $J=3/2$ degree of freedom is even under
inversion so that $P$ acts trivially on the Hilbert space of the local
degrees of freedom and all basis matrices are even. If $P$ is not
trivial, there are also odd basis matrices, which require even
functions of momentum. Hence, in general
$\mathcal{H}_\mathrm{dist}(\mathbf{k})$ need not be odd in
$\mathbf{k}$ but it always contains terms that lead to odd corrections to the quasiparticle energies and can be interpreted as a tilting of the quasiparticle bands. In the following, we only consider the case that $\mathcal{H}_\mathrm{dist}(\mathbf{k})$ is completely odd.

On the other hand, the quasiparticle-band crossings at zero energy are protected by $\mathcal{C}P$ symmetry \cite{KST14,ZSW16,ABT17,BAM18}. When this symmetry is broken the crossings generally gap out. If there were only the gapping at zero energy the situation would be very similar to the BCS theory of superconductivity. However, the tilting of the band structure can shift the gapped out crossings away from zero energy and can in fact lead to residual Fermi surfaces \cite{OhM20}. A similar competition between the energy shifts and gaps occurs for materials that already break inversion symmetry in the normal state~\cite{LiH20}.

Remaining symmetries of the distorted superconductor can prevent the tilting or the splitting on certain high-symmetry elements in the Brillouin zone. The detailed analysis for our model is relegated to Appendix \ref{app.constraints}. In short, we find the following types of symmetry constraints: (a) The distortional contribution to the electronic Hamiltonian can vanish for certain distortions $\mathbf{p}$ and momenta $\mathbf{k}$. This situation can lead to unexpectedly high symmetry, also in combination with the following cases. (b) Spatial twofold symmetry operations, i.e., mirror planes and twofold rotation axes, act like the inversion symmetry on the symmetry element normal to the mirror plane or rotation axis. This protects the symmetry of the energy spectrum and the existence of the $\mathbb{Z}_2$ invariant and thus of band crossings. (c) For certain pairing states, there is no pairing for some bands \emph{within} the mirror plane or on the rotation axis, which protects crossings of these bands. (d) Magnetic twofold symmetry operations, i.e., mirror planes and twofold rotation axes multiplied by time reversal, act like time-reversal symmetry within the mirror plane or on the rotation axis. This protects the symmetry of the energy spectrum but not band crossings.

\begin{table}
\caption{\label{tab.symcon}Symmetry constraints on the quasiparticle band structure for the $T_{2g}$, $\Delta\,(1,i,0)$ pairing state on the caesium-chloride structure.}
\begin{ruledtabular}
\begin{tabular}{cccc}
 & & symmetric & protected \\
\raisebox{1.3ex}[-1.3ex]{distortion $\mathbf{p}$} & \raisebox{1.3ex}[-1.3ex]{momentum $\mathbf{k}$} &
  spectrum & crossings \\ \hline \hline
any & $(0,0,k_z)$ & yes & yes \\
$(p_x,p_y,0)$ & $(k_x,k_y,0)$ & no & some bands \\
$(p_x,0,p_z)$ & $(k_x,0,k_z)$ & yes & no \\
$(0,p_y,p_z)$ & $(0,k_y,k_z)$ & yes & no \\
$(p_x,p_x,p_z)$ & $(k_x,k_x,k_z)$ & yes & no \\
$(p_x,-p_x,p_z)$ & $(k_x,-k_x,k_z)$ & yes & no \\
$(0,0,p_z)$ & $(k_x,k_y,0)$ & yes & yes
\end{tabular}
\end{ruledtabular}
\end{table}

We summarize the constraints for our model in Table \ref{tab.symcon}. Evidently, the quasiparticle bands cannot be completely gapped out for the $T_{2g}$, $\Delta\,(1,i,0)$ pairing state for any local distortion. As an example for how the remaining nodes may look like, we refer to Fig.\ 8 in Ref.\ \cite{TSA17}: Here, band crossings at zero energy are also protected in the $k_xk_y$ plane and on the $k_z$ axis, leading to line and point nodes, respectively. Of course, in Ref.\ \cite{TSA17} inversion symmetry is already broken in the normal state and does not break spontaneously.

\section{Instability of Bogoliubov Fermi surfaces}
\label{sec.BFS_instab}

To assess a possible distortional instability, we have to minimize the free energy as a function of all distortional order parameters $\bPhi=(\Phi_1,\Phi_2,\ldots)$. We here restrict ourselves to the zero-tem\-pe\-ra\-ture limit, in which the free energy becomes the internal energy. We also assume that the system remains in the harmonic regime, which is reasonable for a weak-coupling instability. The internal energy has the form
\begin{equation}
U = U_\mathrm{el}(\Phi_1,\ldots,\Delta) + \frac{1}{2} \sum_\nu K_\nu \Phi_\nu^2 ,
\label{Clog.Uel.3}
\end{equation}
where the first term is the energy of the occupied electronic states, which depends on the distortions and also on the superconducting pairing amplitude $\Delta$, and the second is the harmonic elastic energy. The elastic constants $K_\nu$ for distortions $\Phi_\nu$ belonging to the same multidimensional irrep must be identical. If such a degeneracy applies only to the normal state but is lifted in the superconducting state the difference between the $K_\nu$ is expected to be of order $|\Delta|^2$ and thus small if the superconducting energy scale is small. All $K_\nu$ must be positive since the system in the normal state is undistorted.

In our example, we have three distortional order parameters $\mathbf{p}=(p_x,p_y,p_z)$ and
\begin{equation}
U = U_\mathrm{el}(\mathbf{p},\Delta) + \frac{1}{2}\, K_{E_u} (p_x^2+p_y^2) + \frac{1}{2}\, K_{A_{2u}}\, p_z^2 .
\end{equation}
The two elastic constants are expected to differ little since they are equal for the cubic normal state.

For the characterization of a possible instability, the dependence of $U_\mathrm{el}$ on the distortions $\Phi_\nu$ for small $\Phi_\nu$ is crucial. We derive the electronic energy $U_\mathrm{el}$ within BCS theory, starting from the BCS Hamiltonian
\begin{align}
{\hat H}_\mathrm{BCS} &= \frac{1}{2} \sum_\mathbf{k} \Psi^\dagger_\mathbf{k}
  \mathcal{H}(\mathbf{k}) \Psi^{\phantom{\dagger}}_\mathbf{k}
  + \frac{1}{2} \sum_{\mathbf{k}} \sum_{n=1}^{n_b} \xi_n(\mathbf{k}) \nonumber \\
&\quad{} + \frac{N}{2V_\mathrm{pair}} \sum_m |\Delta_m|^2  ,
\label{Clog.HBCS.2}
\end{align}
where $\mathcal{H}(\mathbf{k})$ is the BdG Hamiltonian of Eq.~(\ref{eq.HBdG.3}),
\begin{equation}
\Psi_\mathbf{k} = \begin{pmatrix}
    c_\mathbf{k} \\ c^\dagger_{-\mathbf{k}}
  \end{pmatrix}
\end{equation}
is a Nambu spinor operator (for our model, $c_\mathbf{k}$ is itself a four-component spinor), $\xi_n(\mathbf{k})$, $n=1,\ldots,n_b$, are the normal-state eigenenergies relative to the chemical potential, and $n_b$ is the dimension of the Hilbert space of the local degrees of freedom, $n_b=4$ for our example. The second term in Eq.\ (\ref{Clog.HBCS.2}) results from anticommuting the fermionic operators to obtain the Nambu form. The last term stems from mean-field decoupling and expresses the fact that the superconducting order parameter has multiple components $\Delta_m$ \cite{ABT17,BAM18}. This term is not relevant for what follows and is omitted from now on. 

In the next step, $\mathcal{H}(\mathbf{k})$ is diagonalized by a Bogoliubov transformation with the new Nambu spinor
\begin{equation}
\Gamma_\mathbf{k} = \begin{pmatrix}
    \gamma_\mathbf{k} \\ \gamma^\dagger_{-\mathbf{k}}
  \end{pmatrix} .
\end{equation}
The result is the BCS Hamiltonian
\begin{align}
{\hat H}_\mathrm{BCS} &= \frac{1}{2} \sum_\mathbf{k} \Gamma^\dagger_\mathbf{k}
  \begin{pmatrix}
    E(\mathbf{k}) & 0 \\
    0 & -E(-\mathbf{k})
  \end{pmatrix} \Gamma^{\phantom{\dagger}}_\mathbf{k}
  + \frac{1}{2} \sum_{\mathbf{k},n} \xi_n(\mathbf{k}) \nonumber \\
&= \sum_{\mathbf{k},n} E_n(\mathbf{k})\, \gamma^\dagger_{\mathbf{k}n} \gamma^{\phantom{\dagger}}_{\mathbf{k}n}
  - \frac{1}{2} \sum_{\mathbf{k},n} E_n(\mathbf{k})
  + \frac{1}{2} \sum_{\mathbf{k},n} \xi_n(\mathbf{k}) ,
\label{Clog.HBCS.4}
\end{align}
where
\begin{equation}
E(\mathbf{k}) = \mathrm{diag}\big(E_1(\mathbf{k}), E_2(\mathbf{k}), \ldots, E_{n_b}(\mathbf{k})\big) .
\end{equation}
The choice of the energies $E_n(\mathbf{k})$ is not unique since one can interchange any pair $E_n(\mathbf{k})$ and $-E_n(-\mathbf{k})$ by a unitary transformation. The results of course do not depend on the choice. If the spectrum at fixed $\mathbf{k}$ is symmetric it is possible to choose all $E_n(\mathbf{k})$ non-negative. However, generally this is not the case.

The electronic internal energy is the ground-state expectation value of ${\hat H}_\mathrm{BCS}$:
\begin{equation}
U_\mathrm{el} = \sum_{\mathbf{k},n} E_n(\mathbf{k})
    \big\langle \gamma^\dagger_{\mathbf{k}n} \gamma^{\phantom{\dagger}}_{\mathbf{k}n}
      \big\rangle_0
  - \frac{1}{2} \sum_{\mathbf{k},n} E_n(\mathbf{k})
  + \frac{1}{2} \sum_{\mathbf{k},n} \xi_n(\mathbf{k}) .
\end{equation}
For $E_n(\mathbf{k})>0$, the BCS ground state $|\mathrm{BCS}\rangle$ satisfies $\gamma_{\mathbf{k}n} |\mathrm{BCS}\rangle = 0$ so that the ground-state expectation value $\langle\gamma^\dagger_{\mathbf{k}n} \gamma^{\phantom{\dagger}}_{\mathbf{k}n}\rangle_0$ vanishes. On the other hand, for $E_n(\mathbf{k})<0$, the expectation value equals unity. We thus obtain~\cite{endnote.normal.limit}
\begin{align}
U_\mathrm{el}
  &= \sum_{\mathbf{k},n} E_n(\mathbf{k})\, \Theta(-E_n(\mathbf{k}))
  - \frac{1}{2} \sum_{\mathbf{k},n} E_n(\mathbf{k})
  + \frac{1}{2} \sum_{\mathbf{k},n} \xi_n(\mathbf{k}) \nonumber \\
&= -\frac{1}{2} \sum_{\mathbf{k}} \sum_{n=1}^{n_b} \big|E_n(\mathbf{k})\big|
  + \frac{1}{2} \sum_{\mathbf{k}} \sum_{n=1}^{n_b} \xi_n(\mathbf{k}) .
\label{Clog.Uel.5}
\end{align}
Note that the first term in the second line contains the sum of the absolute values of one half of the eigenvalues of the BdG Hamiltonian $\mathcal{H}(\mathbf{k})$. For what follows, it is useful to extend the range of $n$ in such a way that
\begin{equation}
E_{n_b+n}(\mathbf{k}) = -E_n(-\mathbf{k}) .
\label{Clog.Enext.3}
\end{equation}
Then, $E_n(\mathbf{k})$ for $n=1,\ldots,2n_b$ are the eigenvalues of $\mathcal{H}(\mathbf{k})$. We can write
\begin{equation}
U_\mathrm{el} = - \frac{1}{4} \sum_{\mathbf{k}} \sum_{n=1}^{2n_b} \big|E_n(\mathbf{k})\big|
  + \frac{1}{2} \sum_{\mathbf{k}} \sum_{n=1}^{n_b} \xi_n(\mathbf{k}) .
\label{Clog.Uel.6}
\end{equation}

An inversion-symmetry-breaking distortion leads to shifts of the normal-state and superconducting-state bands that are odd in momentum $\mathbf{k}$. There are also even-in-mo\-men\-tum shifts but they first occur at second order, $\mathcal{O}(\Phi^2)$, in the distortion and are irrelevant for our analysis and omitted from now on. For the normal-state bands $\xi_n(\mathbf{k})$, the odd-in-momentum shifts clearly drop out of Eq.\ (\ref{Clog.Uel.6}) and can be ignored~\cite{endnote.decoupling.term}.

\subsection{Distortional instability}

The effects of the odd-in-momentum band shifts and the splitting of crossings on the electronic internal energy $U_\mathrm{el}$ can now be calculated. We relegate the derivation to Appendix \ref{app.instability}, where we consider the cases of pure shifts, pure splitting, and the combination of both in turn. We find that the band shifts alone lead to a negative contribution to the internal energy of order $\mathcal{O}(\Phi^2)$ in the distortions. Hence, the prefactor of $\Phi^2$ must be sufficiently large to overcome the positive elastic energy in Eq.~(\ref{Clog.Uel.3}). This would be a strong-coupling instability.

\begin{figure}[t!]
\includegraphics[width=0.8\columnwidth]{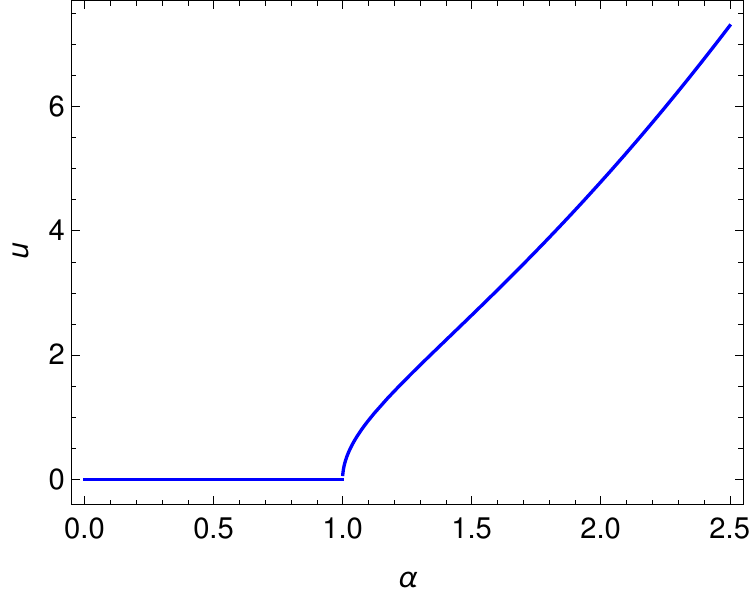}
\caption{\label{fig.uofalpha}Enhancement $u$ of the coefficient of the quadratic term $\mathcal{O}(\Phi^2)$ in the internal energy due to odd-in-momentum band shifts. $\alpha$ is the odd part of the band shift in units of the distortional gap. For $\alpha>1$, residual BFSs exist.}
\end{figure}

The situation changes when the opening of gaps is taken into account. In this case, the internal energy always contains a Cooper logarithm of the form $\Phi^2 \ln \Phi$. For the generic case with both shifts and splittings, the distortional contribution to the electronic internal energy reads as; see Appendix~\ref{app.instability},
\begin{align}
\delta U_\mathrm{el} &= \frac{V}{16\pi^3} \sum_{n=1}^{n_b} \int_\mathrm{BFS}
  d^2k_F\, \frac{\eta_n^2(\mathbf{k}_F)}{v_n(\mathbf{k}_F)}\,
  \bigg[ - \ln 2 - \frac{1}{2} \nonumber \\
&\quad{} + \ln \frac{\eta_0}{v_n(\mathbf{k}_F)\Lambda}
  + u(\alpha_n(\mathbf{k}_F))
  + \ln \frac{\eta_n(\mathbf{k}_F)}{\eta_0} \bigg] ,
\label{Clog.Uel.16}
\end{align}
where $\eta_n(\mathbf{k}_F)$ is half the distortional energy gap of band $n$ at momentum $\mathbf{k}_F$ on the unperturbed BFS, $v_n(\mathbf{k}_F)$ is the magnitude of the Fermi velocity $\mathbf{v}_n(\mathbf{k}_F) \equiv \partial E_n(\mathbf{k})/\partial\mathbf{k}|_{\mathbf{k}=\mathbf{k}_F}$ of the quasiparticles at $\mathbf{k}_F$, $\alpha_n(\mathbf{k}_F) \equiv {|\delta E_n(\mathbf{k}_F)|}/{\eta_n(\mathbf{k}_F)}$ is the band shift in units of the gap (both are of first order in $\Phi$), $\eta_0$ is an arbitrary constant reference energy, and
\begin{equation}
u(\alpha) \equiv \left\{ \begin{array}{ll}
  \displaystyle \alpha \sqrt{\alpha^2-1}
    + \ln\Big(\sqrt{\alpha^2-1}+\alpha\Big) &
  \mbox{for $\alpha>1$,} \\[0.5ex]
  0 & \mbox{for $\alpha\le 1$.}
  \end{array} \right. .
\end{equation}
The function $u(\alpha)$ is plotted in Fig.\ \ref{fig.uofalpha}. The prefactor of the logarithmic term in Eq.\ (\ref{Clog.Uel.16}) is positive and so drives a weak-coupling instability: $\delta U_\mathrm{el}$ can always be reduced by making $\bPhi$ and thus $\eta_n$ nonzero.

We draw a number of conclusions regarding the effect of the odd-in-momentum band shifts, which lead to $\alpha_n(\mathbf{k}_F)>0$. First, where they are smaller than the gap at the Fermi energy they do not have any effect on the weak-coupling instability. Second, where they are larger---and residual BFSs exist---they do not affect the Cooper logarithm but only renormalize the prefactor of the quadratic term $\mathcal{O}(\Phi^2)$. Larger shifts lead to a positive correction, which is added to the positive elastic energy. They thus increase the effective stiffness of the crystal and reduce the distortion at zero temperature, as expected. However, as long as there is any band splitting anywhere on the unperturbed BFSs, the Cooper logarithm and the weak-coupling instability survive. This is true even if the band shifts are larger than the splitting \emph{everywhere} so that the BFSs are nowhere gapped out. In this situation, the BFSs are split into two surfaces, one inside the other.

\subsection{Which distortion is stabilized?}

So far, we have found that a distortional weak-coupling instability always exists. Typically, the distortional order parameter $\bPhi$ has multiple components. The instability will select those distortions that minimize the full internal energy $U = U_\mathrm{el} + (1/2) \sum_\nu K_\nu \Phi_\nu^2$. Equation (\ref{Clog.Uel.16}) shows that  distortions are favored that lead to the largest splitting, or more precisely the largest $\eta_n^2(\mathbf{k}_F)/v_n(\mathbf{k}_F)$, averaged over the BFSs for a given magnitude $\Phi$. In the following, we restrict ourselves to our model.

Since the distortion is small at a weak-coupling instability we can expand the band splittings,
\begin{equation}
\eta_n(\mathbf{k}_F) \cong \left. \frac{\partial \eta_n(\mathbf{k}_F)}
  {\partial\mathbf{p}} \right|_{\mathbf{p}=0} \cdot \mathbf{p}
  \equiv \mathbf{c}_n(\mathbf{k}_F) \cdot \mathbf{p} .
\end{equation}
The logarithm is then
\begin{equation}
\ln \frac{\eta_n(\mathbf{k}_F)}{\eta_0}
  \cong \ln \frac{\mathbf{c}_n(\mathbf{k}_F) \cdot \mathbf{p}}{\eta_0}
  = \ln \frac{\mathbf{c}_n(\mathbf{k}_F) \cdot p_0\hat{\mathbf{p}}}{\eta_0}
  + \ln \frac{p}{p_0} ,
\end{equation}
where $p=|\mathbf{p}|$ and $p_0$ is an arbitrary constant with the same dimension as the distortional order parameter. The change in the internal energy reads as
\begin{align}
\delta U &\cong \frac{V}{16\pi^3} \sum_{n=1}^{n_b} \int_\mathrm{BFS}
  d^2k_F\, \frac{(\mathbf{c}_n(\mathbf{k}_F) \cdot \mathbf{p})^2}
  {v_n(\mathbf{k}_F)}\, \bigg[ {-} \ln 2 - \frac{1}{2} \nonumber \\
&\quad{} + \ln \frac{\mathbf{c}_n(\mathbf{k}_F) \cdot p_0\hat{\mathbf{p}}}
    {v_n(\mathbf{k}_F)\Lambda}
  + u(\alpha_n(\mathbf{k}_F))
  + \ln \frac{p}{p_0} \bigg] \nonumber \\
&\quad{}+ \frac{1}{2}\, K_{E_u}\, (p_x^2+p_y^2)
  + \frac{1}{2}\, K_{A_{2u}}\, p_z^2 .
\end{align}
We assume that the crystal is far from a distortional instability in the normal state, i.e., that the elastic internal energy is large. We can then neglect the renormalization of the $\mathcal{O}(p^2)$ term by electronic contributions, leading to
\begin{align}
\delta U &\cong \frac{V}{16\pi^3} \sum_{n=1}^{n_b} \int_\mathrm{BFS}
  d^2k_F\, \frac{(\mathbf{c}_n(\mathbf{k}_F) \cdot \mathbf{p})^2}
    {v_n(\mathbf{k}_F)}\, \ln \frac{p}{p_0} \nonumber \\
&\quad{} + \frac{1}{2}\, K_{E_u} (p_x^2+p_y^2)
  + \frac{1}{2}\, K_{A_{2u}} p_z^2 \nonumber \\
&= \frac{1}{2} \sum_{ij} C_{ij}\, p_i p_j \ln \frac{p}{p_0}
  + \frac{1}{2}\, K_{E_u} (p_x^2+p_y^2)
  + \frac{1}{2}\, K_{A_{2u}} p_z^2 ,
\end{align}
with
\begin{align}
C_{ij} &\equiv \frac{V}{8\pi^3} \sum_{n=1}^{n_b} \int_\mathrm{BFS} \!
  d^2k_F\, \frac{c_{n,i}(\mathbf{k}_F)\, c_{n,j}(\mathbf{k}_F)}
    {v_n(\mathbf{k}_F)} \nonumber \\
&= \frac{V}{8\pi^3} \sum_{n=1}^{n_b} \int_\mathrm{BFS} \!
  d^2k_F\, \frac{1}{v_n(\mathbf{k}_F)} \nonumber \\
&\quad{} \times \left. \frac{\partial \eta_n(\mathbf{k}_F)}{\partial p_i} \right|_{\mathbf{p}=0}
  \left. \frac{\partial \eta_n(\mathbf{k}_F)}{\partial p_j} \right|_{\mathbf{p}=0} .
\label{Clog.Cdef.3}
\end{align}
We find that distortions are favored which maximize $C_{ij}$.

Now $(1/2) \sum_{ij} C_{ij}\, p_i p_j$ must have the full $4/mm'm'$ symmetry of the superconducting state when $\mathbf{p}$ is also transformed. In other words, the bilinear form must transform according to $A_{1g}$. As noted above, the doublet $(p_x,p_y)$ belongs to $E_u$ and $p_z$ belongs to $A_{2u}$. Since we have expanded to second order, the matrix $C$ does not depend on $\mathbf{p}$. Hence, all coefficients are constants, which transform according to $A_{1g}$. The only allowed components are then $C_{xx}=C_{yy} \equiv C_{E_u}$ and $C_{zz} \equiv C_{A_{2u}}$. We end up with
\begin{align}
\delta U &\cong \frac{1}{2}\, \big[ C_{E_u} (p_x^2 + p_y^2) + C_{A_{2u}} p_z^2 \big] \, \ln \frac{p}{p_0}
  \nonumber \\
&\quad{} + \frac{1}{2}\, \big[ K_{E_u} (p_x^2+p_y^2)  + K_{A_{2u}} p_z^2 \big] .
\end{align}
The symmetry of $\delta U$ is in fact much higher than the expected $4/mm'm'$: this group contains a fourfold rotation axis parallel to $\hat{\mathbf{z}}$ but $\delta U$ has continuous $\mathrm{SO}(2)$ rotation symmetry about the \textit{z} axis. This is due to the well-known effect that quadratic forms with rotation symmetry of order 3 or higher have continuous rotation symmetry. The logarithmic factor has continuous symmetry in any case.

The energy change $\delta U$ has a local maximum at $\mathbf{p}=0$. Minimization can lead to $\mathbf{p}=(p\cos\varphi,p\sin\varphi,0)$ with arbitrary angle $\varphi$ (easy plane) or to $\mathbf{p}=(0,0,p)$ (easy axis). For in-plane distortions, we find
\begin{equation}
0 = \frac{\partial}{\partial p}\, \delta U
  \cong C_{E_u}\, p\, \ln \frac{p}{p_0}
    + \frac{1}{2}\, C_{E_u}\, p + K_{E_u}\, p ,
\end{equation}
which with $p\neq 0$ implies
\begin{equation}
p \cong \frac{p_0}{\sqrt{e}} \exp\left( - \frac{K_{E_u}}{C_{E_u}} \right) .
\end{equation}
The change in the internal energy is then
\begin{align}
\delta U &\cong - \frac{C_{E_u}p_0^2}{2e}
  \left( \frac{K_{E_u}}{C_{E_u}} + \frac{1}{2} \right)
  \exp\left( - \frac{2K_{E_u}}{C_{E_u}} \right) \nonumber \\
&\quad{} + \frac{K_{E_u}p_0^2}{2e}\, \exp\left( - \frac{2K_{E_u}}{C_{E_u}} \right)
  \nonumber \\
&= - \frac{C_{E_u}p_0^2}{4e}\, \exp\left( - \frac{2K_{E_u}}{C_{E_u}} \right) .
\end{align}
Analogously, for distortions along the \textit{z} axis, we obtain
\begin{align}
p &\cong \frac{p_0}{\sqrt{e}}
  \exp\left( - \frac{K_{A_{2u}}}{C_{A_{2u}}} \right) , \\
\delta U &\cong - \frac{C_{A_{2u}}p_0^2}{4e}\,
  \exp\left( - \frac{2K_{A_{2u}}}{C_{A_{2u}}} \right) .
\end{align}
As discussed above, the stiffness constants $K_{E_u}$ and $K_{A_{2u}}$ are expected to be similar since they are equal in the normal state. On the other hand, the parameters $C_{E_u}$ and $C_{A_{2u}}$ describe the gapping of the quasiparticle bands of the anisotropic superconducting state due to distortions and have no reason to be similar. Hence, the difference in internal energy for in-plane and \textit{z} axis distortions should be dominated by the difference between $C_{E_u}$ and $C_{A_{2u}}$. Replacing $K_{E_u} \approx K_{A_{2u}}$ by $K$, the energy change
\begin{equation}
\delta U(C) \cong - \frac{C\, p_0^2}{4e}\,
  \exp\left( - \frac{2K}{C} \right)
\end{equation}
is negative for all $C>0$ and is a monotonically decreasing function of $C$. Hence, distortions with larger $C$ are favored but the energy gain is exponentially small for small $C$, typical for a weak-coupling instability.

We now address the question how strongly the internal energy differentiates between different distortions. Since $\delta U(C)$ depends very strongly on $C$ we expect that modest differences in $C$ lead to large differences in energy and thus, following BCS theory, to large differences in the distortional critical temperature. The relative difference between the energy gains for in-plane and \textit{z} axis distortions is
\begin{align}
&2\, \frac{\delta U(C_{E_u}) - \delta U(C_{A_{2u}})}{\delta U(C_{E_u}) + \delta U(C_{A_{2u}})} \nonumber \\
  &= -2\, \frac{C_{E_u}\, e^{-2K/C_{E_u}} - C_{A_{2u}}\, e^{-2K/C_{A_{2u}}}}
    {C_{E_u}\, e^{-2K/C_{E_u}} + C_{A_{2u}}\, e^{-2K/C_{A_{2u}}}} \nonumber \\
&= -2\, \tanh \left( -\frac{K}{C_{E_u}} + \frac{K}{C_{A_{2u}}} + \ln\frac{C_{E_u}}{C_{A_{2u}}} \right) .
\end{align}
In the weak-coupling limit, $K/C_{E_u}$ and $K/C_{A_{2u}}$ are large. Since $C_{E_u}$ and $C_{A_{2u}}$ are not expected to be similar the first two terms in the argument of the hyperbolic tangent together are large in magnitude. The logarithm generically does not compensate for this. Consequently, the argument is large in magnitude and the whole expression is close to $-2 \sgn (C_{E_u} - C_{A_{2u}})$. Hence, the favored distortion typically is hugely favored.

\subsection{Numerical results}
\label{sub.numerics}

In the following, we illustrate the measure $C_{ij}$ defined in Eq.\ (\ref{Clog.Cdef.3}) by numerical results. The undistorted normal-state Hamiltonian $H_N$ in Eq.\ (\ref{eq.HN.3}) contains functions $c_n(\mathbf{k})$, which we choose in accordance with point-group and translation symmetries as
\begin{align}
c_0(\mathbf{k}) &= -2t_0\, (\cos k_x + \cos k_y + \cos k_z) - \mu , \\
c_1(\mathbf{k}) &= 4t_1\, \sin k_y \sin k_z , \\
c_2(\mathbf{k}) &= 4t_1\, \sin k_z \sin k_x , \\
c_3(\mathbf{k}) &= 4t_1\, \sin k_x \sin k_y , \\
c_4(\mathbf{k}) &= 4t_2\, (-\cos k_x + \cos k_y) \cos k_z , \\
c_5(\mathbf{k}) &= \frac{4t_2}{\sqrt{3}}\, ( 2 \cos k_x \cos k_y - \cos k_y \cos k_z \nonumber \\
&\quad{}- \cos k_z \cos k_x ) .
\end{align}
We take $t_0=2$, $t_1=-1.5$, $t_2=-1.7$, and $\mu=-12.6$. The normal state then has a small hole-type Fermi surface in the lower band, which to good approximation belongs to the magnetic quantum numbers $\pm 3/2$ with respect to the momentum-dependent spin quantization axis. A large value of $\Delta=0.5/\sqrt{2}$ is chosen for the superconducting gap amplitude in the order parameter $\Delta\,(1,i,0)$ to obtain sizable BFSs. The qualitative results do not depend on the amplitude.

One problem here is the large number of allowed coupling terms between the electronic system and distortions appearing in Eqs.\ (\ref{eq.Hdist.pz}) and (\ref{eq.Hdist.pxpy}). To reduce the complexity, we (a) neglect couplings that require superconductivity (with basis functions $e^l_\Gamma$), (b) assume the simplest trigonometric form of basis functions that is consistent with the caesium chloride structure (i.e., with the simple cubic Bravais lattice), and (c) consider each of the remaining terms in Eqs.\ (\ref{eq.Hdist.pz}) and (\ref{eq.Hdist.pxpy}) separately. Table \ref{tab.CC} shows numerical results for $C_{E_u}/V$ and $C_{A_{2u}}/V$ as defined in Eq.\ (\ref{Clog.Cdef.3}). Note that the numerics show that the terms appearing in $H_{p_z}(\mathbf{k})$ [Eq.\ (\ref{eq.Hdist.pz})] indeed only favor distortions along the \textit{z} direction, whereas terms in $H_{p_x,p_y}(\mathbf{k})$ [Eq.\ (\ref{eq.Hdist.pxpy})] only favor in-plane distortions. In a real system, all terms will be present simultaneously since they are all allowed by symmetry. We observe that the contributions with form factors $d^l_\Gamma(\mathbf{k})$ belonging to $A_{1u}$ and $E_u$ are the only sizable ones. This is a consequence of having a small normal-state Fermi surface: all other form factors are suppressed by being of higher order in momentum.

\begin{table}
\caption{\label{tab.CC}The parameters $C_{E_u}/V$ and $C_{A_{2u}}/V$ that determine the energy gain due to various allowed distortion terms, labeled by the form factors in Eqs.\ (\ref{eq.Hdist.pz}) and (\ref{eq.Hdist.pxpy}). For each of them, the simplest trigonometric form consistent with the simple cubic Bravais lattice has been assumed and its prefactor has been set to unity. The basis matrices $h$ in spin space have been normalized so that $\Tr h^2=4$; see Table \ref{tab.basis}. Large $C_{E_u}/V$ favors in-plane distortions, whereas large $C_{A_{2u}}/V$ favors distortions along the \textit{z} axis. Entries ``$0$'' mean zero by symmetry, confirmed by the numerics, while ``$0.00$'' signifies a small positive number that is rounded to zero.}
\begin{ruledtabular}
\begin{tabular}{ccc}
 & $C_{E_u}/V$ ($10^{-5}$) & $C_{A_{2u}}/V$ ($10^{-5}$) \\ \hline\hline
$d^1_{A_{1u}}$ & $5.23$ & $0$ \\
$d^2_{A_{1u}}$ & $8.57$ & $0$ \\
$d^3_{A_{1u}}$ & $27.1$ & $0$ \\
$d^1_{A_{2u}}$ & $0$ & $0.00$ \\
$d^2_{A_{2u}}$ & $0$ & $0.00$ \\
$d^3_{A_{2u}}$ & $0.00$ & $0$ \\
$d^4_{A_{2u}}$ & $0.00$ & $0$ \\
$d^5_{A_{2u}}$ & $0.00$ & $0$ \\
$d^1_{B_{1u}}$ & $0$ & $0.00$ \\
$d^2_{B_{1u}}$ & $0.09$ & $0$ \\
$d^3_{B_{1u}}$ & $0.10$ & $0$ \\
$d^4_{B_{1u}}$ & $0.05$ & $0$ \\
$d^1_{B_{2u}}$ & $0$ & $0.00$ \\
$d^2_{B_{2u}}$ & $0.13$ & $0$ \\
$d^3_{B_{2u}}$ & $0.12$ & $0$ \\
$d^4_{B_{2u}}$ & $0.12$ & $0$ \\
$d^1_{E_u}$ & $0$ & $1.16$ \\
$d^2_{E_u}$ & $0$ & $20.2$ \\
$d^3_{E_u}$ & $0$ & $449$ \\
$d^4_{E_u}$ & $8.1$ & $0$ \\
$d^5_{E_u}$ & $5.44$ & $0$ \\
$d^6_{E_u}$ & $8.98$ & $0$ \\
$d^7_{E_u}$ & $8.24$ & $0$
\end{tabular}
\end{ruledtabular}
\end{table}

Among the $A_{1u}$ and $E_u$ contributions, only three out of ten favor distortions along the \textit{z} direction but the largest entry by more than one order of magnitude does so. The dominant entry refers to the term
\begin{equation}
\vec d_{E_u}^{\:3}(\mathbf{k}) \cdot \begin{pmatrix} J_x(J_y^2-J_z^2) + (J_y^2-J_z^2)J_x \\[0.5ex]
  - J_y(J_z^2-J_x^2) - (J_z^2-J_x^2)J_y \end{pmatrix}
\label{eq.domd.3}
\end{equation}
in $H_{p_z}(\mathbf{k})$. The $E_u$ prefactors are $\vec d_{E_u}^{\:l}(\mathbf{k}) = (\sin k_y, -\sin k_x)$ for all $l=1,\ldots,7$. Why is this contribution so much more favorable compared to the simpler ones, for example the term involving $(J_x,J_y)$? While the answer certainly depends on details of the model, one can understand the relative importance of terms based on the pseudospin picture, as we show at the end of the next section.

\section{Pseudospin picture}
\label{sec.pseudo}

In our model, the normal-state bands are twofold degenerate at general $\mathbf{k}$ because of time-reversal and inversion symmetries. This twofold degeneracy can be pa\-ra\-me\-trized by a pseudospin of length $1/2$. If the normal-state bands are sufficiently far apart on the superconducting energy scale the effect of interband pairing can be treated perturbatively. Such a treatment leads to (a) a spin-independent shift $\gamma(\mathbf{k})$ of the normal-state band energies, (b) a pseudomagnetic field $\mathbf{h}(\mathbf{k})$ coupling to the pseudospin, and (c) an intraband pseudospin-singlet pairing term with an amplitude $\psi(\mathbf{k})$ that encodes the uninflated nodes \cite{ABT17,BAM18}. Since the intraband pairing is of pseudo-spin-singlet nature, the momentum-dependent amplitude $\psi(\mathbf{k})$ must carry the full symmetry information of the pairing state in the original multiband system. Here, we discuss the distortional correction $H_\text{dist}(\mathbf{k})$ and its consequences in the pseudospin picture.

\begin{table}[th]
\caption{\label{tab.pseudo.basis}Basis matrices of the two-dimensional pseudospin Hilbert space. The last two columns show the corresponding irreps for the magnetic point groups $m\bar{3}m1'$ and $4/mm'm'$. The correct order and signs of the $E_g$ doublet for $4/mm'm'$, again using $(xz,yz)$ as the template, are $(\sigma_1,\sigma_2)$.}
\begin{ruledtabular}
\begin{tabular}{ccc}
 & \multicolumn{2}{c}{irrep} \\[-0.7ex]
\raisebox{1.8ex}[-1.8ex]{basis matrix} & \rule[-1ex]{0ex}{4ex} $m\bar{3}m1'$ & $4/mm'm'$ \\ \hline\hline
$\sigma_0$ & $A_{1g+}$ & $A_{1g}$ \\ \hline
$\sigma_1$ & & \\
$\sigma_2$ & $T_{1g-}$ & \raisebox{1.3ex}[-1.3ex]{$E_g$} \\ \cline{3-3}
$\sigma_3$ & & $A_{1g}$
\end{tabular}
\end{ruledtabular}
\end{table}

The correction to the electronic Hamiltonian due to uniform distortions, $H_\text{dist}(\mathbf{k})$, is constrained by symmetry. It must be invariant under all point-group operations if the distortional order parameters {\boldmath$\Phi$} are transformed simultaneously. Moreover, $H_\text{dist}(\mathbf{k})$ is linear in {\boldmath$\Phi$} in the harmonic limit. The normal-state Hamiltonian is of course parametrized by the $2\times 2$ identity matrix $\sigma_0\equiv\boldone$ and the three Pauli matrices $\sigma_1$, $\sigma_2$, $\sigma_3$. For our caesium chloride model, the symmetry classification of these basis matrices is given in Table \ref{tab.pseudo.basis}, which corresponds to Table \ref{tab.basis} for the full model.

As noted above, the distortion modes $\mathbf{p}$ transform according to $T_{1u+}$ of $m\bar{3}m1'$ in the normal state. We now only need the reductions $T_{1u+} \otimes A_{1g+} = T_{1u+}$, which does not have any momentum basis functions and thus does not lead to an allowed term in $H_\text{dist}(\mathbf{k})$, and $T_{1u+} \otimes T_{1g-} = A_{1u-} \oplus E_{u-} \oplus T_{1u-} \oplus T_{2u-}$. Hence, in the normal state only $\sigma_1$, $\sigma_2$, $\sigma_3$ can appear in $H_\text{dist}(\mathbf{k})$. To get an idea of the leading terms, we ignore the reduction of the point group from $m\bar{3}m1'$ to $4/mm'm'$ due to superconductivity.

The identity $T_{1u+} \otimes T_{1g-} = A_{1u-} \oplus E_{u-} \oplus T_{1u-} \oplus T_{2u-}$ and a table of $O_h$ basis functions show that the electronic Hamiltonian due to the distortion must have the form
\begin{align}
H_\text{dist}(\mathbf{k}) &= d_{A_{1u-}}(\mathbf{k})\, \mathbf{p} \cdot \bsigma \nonumber \\
&\quad{} + {\vec d}_{E_{u-}}(\mathbf{k})
    \cdot \begin{pmatrix} p_x \sigma_1 - p_y \sigma_2  \\
       (2p_z \sigma_3 - p_x \sigma_1 - p_y \sigma_2)/\sqrt{3} \end{pmatrix} \nonumber \\
&\quad{} + {\vec d}_{T_{1u-}}(\mathbf{k})
    \cdot \begin{pmatrix} p_y \sigma_3 - p_z \sigma_2 \\
      p_z \sigma_1 - p_x \sigma_3 \\
      p_x \sigma_2 - p_y \sigma_1 \end{pmatrix} \nonumber \\
&\quad{} + {\vec d}_{T_{2u-}}(\mathbf{k})
    \cdot \begin{pmatrix} p_y \sigma_3 + p_z \sigma_2 \\
      p_z \sigma_1 + p_x \sigma_3 \\
      p_x \sigma_2 + p_y \sigma_1 \end{pmatrix}
  \equiv \mathbf{g}(\mathbf{k}) \cdot \bsigma ,
\end{align}
where we have used the center dot for two-dimensional and three-dimensional scalar products indiscriminately. This result is quite interesting. For example, the first term has the same form as the coupling of the pseudomagnetic field. The direction is fixed by the distortion $\mathbf{p}$ but it has a very complicated form in momentum space since the lowest-order polynomial basis function of $A_{1u-}$ is $k_xk_yk_z\, [k_x^4\, (k_y^2-k_z^2) + k_y^4\, (k_z^2-k_x^2) + k_z^4\, (k_x^2-k_y^2)]$. One generally has the prejudice that lower-order basis functions dominate. If we hence include only the basis functions of lowest order, which belong to $T_{1u-}$, we obtain
\begin{align}
H_\text{dist}(\mathbf{k}) &\cong
  \alpha\, \mathbf{k} \cdot \begin{pmatrix} p_y \sigma_3 - p_z \sigma_2 \\
      p_z \sigma_1 - p_x \sigma_3 \\
      p_x \sigma_2 - p_y \sigma_1 \end{pmatrix}
  = \alpha\, \mathbf{k} \cdot ( \mathbf{p} \times \bsigma ) \nonumber \\
&= -\alpha\, ( \mathbf{p} \times \mathbf{k} ) \cdot \bsigma
\end{align}
and thus
\begin{equation}
\mathbf{g}(\mathbf{k}) \cong -\alpha\, \mathbf{p} \times \mathbf{k} ,
\label{pseudo.g.3}
\end{equation}
where $\alpha$ is a constant. This has the form of an antisymmetric spin-orbit coupling (ASOC)
of Rashba type with the role of the electric field played by the distortional order parameter $\mathbf{p}$. This is plausible since the distortion in our model is accompanied by a dipole moment.

In the pseudospin picture, the BdG Hamiltonian has the general form
\begin{equation}
\mathcal{H}(\mathbf{k}) = \begin{pmatrix}
    H_N(\mathbf{k}) & \psi(\mathbf{k})\, i\sigma_2 \\
    -\psi^*(\mathbf{k})\, i\sigma_2 & - H_N^T(-\mathbf{k})
  \end{pmatrix} ,
\end{equation}
with
\begin{equation}
H_N(\mathbf{k}) = \xi(\mathbf{k})\, \sigma_0 + \mathbf{g}(\mathbf{k}) \cdot \bsigma
   + \mathbf{h}(\mathbf{k}) \cdot \bsigma ,
\end{equation}
where $\xi(\mathbf{k}) \equiv \epsilon(\mathbf{k}) - \mu$ is the normal-state dispersion including the chemical potential. The spin-independent renormalization of the dispersion due to interband pairing has been absorbed into $\xi(\mathbf{k})$. The ASOC and the pseudomagnetic-field term look quite similar but $\mathbf{g}(\mathbf{k})$ is odd in $\mathbf{k}$, whereas $\mathbf{h}(\mathbf{k})$ is even. We now use the behavior under inversion and suppress the common argument $\mathbf{k}$, which yields
\begin{equation}
\mathcal{H} = \begin{pmatrix}
    \xi \sigma_0 + \mathbf{g} \cdot \bsigma
      + \mathbf{h} \cdot \bsigma & \psi\, i\sigma_2 \\
    -\psi^*\, i\sigma_2 & -\xi \sigma_0
    + \mathbf{g} \cdot \bsigma^T - \mathbf{h} \cdot \bsigma^T
  \end{pmatrix} .
\end{equation}
The BdG Hamiltonian only contains the two vectors $\mathbf{h}$ and $\mathbf{g}$. Since we are only interested in the eigenvalues we can simplify the problem by choosing the coordinate system in spin space in such a way that $\mathbf{h}$ is along $\hat{\mathbf{z}}$ and $\mathbf{g}$ is in the \textit{xz} plane. Moreover, the eigenvalues cannot depend on the phase of $\psi$ so that we can choose a (generally momentum-dependent) gauge that makes $\psi$ real. This gives
\begin{widetext}
\begin{equation}
\mathcal{H} = \begin{pmatrix}
    \xi + g_z + h & g_x & 0 & \psi \\
    g_x & \xi - g_z - h & -\psi & 0 \\
    0 & -\psi & -\xi + g_z - h & g_x \\
    \psi & 0 & g_x & -\xi - g_z + h
  \end{pmatrix} .
\label{pseudo.Hsimple.3}
\end{equation}
It is obvious that the Hamiltonian consists of decoupled blocks for $\psi=0$. This reflects that there is no superconducting gap for $\psi=0$ and thus that bands can cross. By interchanging the second and fourth basis states, corresponding to a particle-hole transformation in the spin-down channel, we rewrite the Hamiltonian as
\begin{equation}
\hat{\mathcal{H}} \equiv \begin{pmatrix}
    h + g_z + \xi & \psi & 0 & g_x \\
    \psi & h - g_z - \xi & g_x & 0 \\
    0 & g_x & -h + g_z - \xi & -\psi \\
    g_x & 0 & -\psi & -h - g_z + \xi
  \end{pmatrix} .
\label{pseudo.Hsimple.5}
\end{equation}
\end{widetext}
Compared to $\mathcal{H}$, $\xi$ and $h$ as well as $\psi$ and $g_x$ are interchanged. However, $\psi$ is even in momentum, which is reflected by $-\psi$ appearing in the lower right block, whereas $g_x$ is odd. We will say that $\hat{\mathcal{H}}$ describes the \emph{dual} model.

The undistorted superconducting state ($g_x=g_z=0$) corresponds, in the dual model, to a \emph{normal} metal with Hamiltonian $\hat H_N = h\, \sigma_0 + \mbox{\boldmath$\gamma$} \cdot \bsigma$, where $\mbox{\boldmath$\gamma$} = (\psi,0,\xi)$. The dual model is indeed metallic---its Fermi surface is the BFS of the original model. The transverse distortional instability ($g_x$) maps onto a superconducting instability. The dual superconducting gap matrix is $g_x \sigma_1 = g_x\, \sigma_3\, i\sigma_2$ so that this is a triplet superconductor with spin along the \textit{z} direction. $g_x$ is odd in momentum, consistent with triplet superconductivity for an inversion-symmetric normal state.

The order parameter $g_z$ is added to the dual magnetic field $\xi$ along the \textit{z} direction. At fixed momentum, this does not change any symmetries and thus does not change the properties of the spectrum. In the original model, $g_z$ renormalizes the dispersion $\xi$. Since $\xi$ changes rapidly in momentum space (the band width is large compared to $g_z$) this is a small effect and leads to a small shift of the normal-state Fermi surface. Since $g_z$ is odd in momentum the band shift breaks the inversion symmetry of the bands and of the normal-state Fermi surface. Since $g_z$ and $-g_z$ are related by a symmetry of the undistorted system, namely by inversion, the effect on the free energy is an even function of $g_z$. Generically, the effect will scale with $g_z^2$ for small $g_z$ and thus at most lead to a strong-coupling instability. This is the Pomeranchuk instability discussed in Ref.~\cite{TIH20}.

On the other hand, the dual superconducting order parameter $g_x$ generically opens a gap, which leads to a Cooper log in the internal energy. It therefore causes a weak-coupling instability \cite{TIH20}. $g_x$ also breaks inversion symmetry and thus together with $g_z$ causes shifts of quasiparticle bands that are odd in momentum. The corresponding change in the free energy is expected to be quadratic and thus changes the energetics but cannot compensate for the Cooper log.

Going back to the original coordinate system, we conclude that the component of the distortion-induced ASOC $\mathbf{g}$ that is parallel to the interband-pairing-induced pseudomagnetic field $\mathbf{h}$ does not cause a weak-coupling instability of the BFSs. On the other hand, the components normal to $\mathbf{h}$ do lead to a weak-coupling instability of the BFSs, which is analogous to the one in BCS theory by means of a duality mapping. These results show that $\mathbf{g}(\mathbf{k})$ with large components normal to $\mathbf{h}(\mathbf{k})$ are favored.

We can now partially understand the numerical results shown in Table \ref{tab.CC}. The pseudomagnetic field $\mathbf{h}(\mathbf{k})$ for the $\Delta\,(1,i,0)$ pairing state is obtained by projecting the time-reversal-odd part of the gap product,
\begin{equation}
\Delta \Delta^\dagger - U_T\, (\Delta \Delta^\dagger)^* U_T^\dagger = \frac{4}{3}\, \Delta^2\, (7J_z - 4J_z^3)
\end{equation}
into the relevant band \cite{BAM18}. Here,
\begin{equation}
U_T = \begin{pmatrix}
    0 & 0 & 0 & 1 \\
    0 & 0 & -1 & 0 \\
    0 & 1 & 0 & 0 \\
    -1 & 0 & 0 & 0
  \end{pmatrix}
\end{equation}
is the unitary part of the time-reversal operator. The result can be written in the form $\mathbf{h}(\mathbf{k}) \cdot \bsigma$. The pairing state is a pseudospin ferromagnet in that $\mathbf{h}(\mathbf{k})$ has a nonzero average. To get an idea about the favored distortion, we can treat $\mathbf{h}(\mathbf{k})$ to leading order; i.e., we replace it by its average, which is parallel to the \textit{z} axis. This implies that $\mathbf{g}(\mathbf{k})$ lying in the \textit{xy} plane is favored. Since $\mathbf{g}(\mathbf{k}) \cong -\alpha\, \mathbf{p} \times \mathbf{k}$ we conclude that $\mathbf{p} \parallel \hat{\mathbf{z}} \parallel \mathbf{h}$ is favored. This agrees with the numerical results of Sec.~\ref{sub.numerics}.

To understand the relative strength of the three distortion terms favoring $\mathbf{p}$ in the \textit{z} direction, we consider the corresponding terms in $H_{p_z}(\mathbf{k})$. For small $\mathbf{k}$, these terms read as
\begin{align}
H_\textrm{dist}^1(\mathbf{k}) &= \frac{2}{\sqrt{5}}\, ( k_y J_x - k_x J_y )\, p_z , \\
H_\textrm{dist}^2(\mathbf{k}) &= \frac{8}{\sqrt{365}}\, ( k_y J_x^3 - k_x J_y^3 )\, p_z , \\
H_\textrm{dist}^3(\mathbf{k}) &= \frac{1}{\sqrt{3}}\, \big( k_y\, [J_x(J_y^2-J_z^2)
  + (J_y^2-J_z^2)J_x] \nonumber \\
&\quad{} + k_x\, [J_y(J_z^2-J_x^2) + (J_z^2-J_x^2)J_y] \big)\, p_z .
\end{align}
After projection into the relevant band, they can be written as $\mathbf{g}^l(\mathbf{k}) \cdot \bsigma$, $l=1,2,3$, which defines their contributions $\mathbf{g}^l(\mathbf{k})$ to the distortion-induced ASOC in the pseudospin picture. The model used for the numerics has a small and roughly spherical normal-state Fermi surface. It is thus justified to employ a spherical approximation for the undistorted Hamiltonian,
\begin{equation}
H_N(\mathbf{k}) \cong (\alpha k^2 - \mu) \boldone + \beta (\mathbf{k} \cdot \mathbf{J})^2 ,
\end{equation}
which has the advantage that all evaluations can be performed analytically. The relevant band is the one with eigenvalues $\pm 3/2$ of $\hat{\mathbf{k}} \cdot \mathbf{J}$, where $\hat{\mathbf{k}} = \mathbf{k}/k$.

We have seen that distortions are favored when the component of the effective ASOC vector $\mathbf{g}^l(\mathbf{k})$ normal to $\mathbf{h}(\mathbf{k})$ is large. A good measure for this is $|\mathbf{g}^l \times \mathbf{h}|^2 = (\mathbf{g}^l\times\mathbf{h})\cdot(\mathbf{g}^l\times\mathbf{h})$. Note that this quantity is independent of the choice of basis for the pseudospin. For the three terms of interest, we obtain
\begin{align}
|\mathbf{g}^1 \times \mathbf{h}|^2 &= 0 , \\
|\mathbf{g}^2 \times \mathbf{h}|^2
  &= \frac{9}{1460}\, k^2 p_z^2 \sin^2\theta\, \Big(4\, [\sin^6\theta + \cos^6\theta ] \nonumber \\
&\qquad{} \times \left[(3 \cos 2 \theta + 5) \cos 4 \phi + 6 \sin^2\theta \right]^2 \nonumber \\
&\quad{} + [6 \cos 2 \theta + 3 \cos 4 \theta + 7]^2 \sin^2 4 \phi \Big) , \\
|\mathbf{g}^3 \times \mathbf{h}|^2
  &= \frac{3}{64}\, k^2 p_z^2 \sin^2\theta\, \Big(4\, [\sin^6\theta + \cos^6 \theta ] \nonumber \\
&\qquad{} \times \left[(3 \cos 2 \theta + 5) \cos 4 \phi - 10 \sin^2\theta \right]^2 \nonumber \\
&\quad{} + [6 \cos 2 \theta +3 \cos 4 \theta + 7]^2 \sin ^2 4 \phi \Big) .
\end{align}
In fact, the full vector $\mathbf{g}^1(\mathbf{k})$ vanishes, not just its components normal to $\mathbf{h}(\mathbf{k})$. This follows from the fact that the operator $H_\textrm{dist}^1(\mathbf{k})$ is proportional to a spin component perpendicular to $\mathbf{k}$, which has vanishing matrix elements between the two states in the relevant band, which are eigenstates of $\hat{\mathbf{k}}\cdot\mathbf{J}$ with eigenvalues $\pm 3/2$. This explains why the entry for $d^1_{E_u}$ in Table \ref{tab.CC} is small (it is not zero because the real model is not spherical).

\begin{figure}[t!]
\includegraphics[width=\columnwidth]{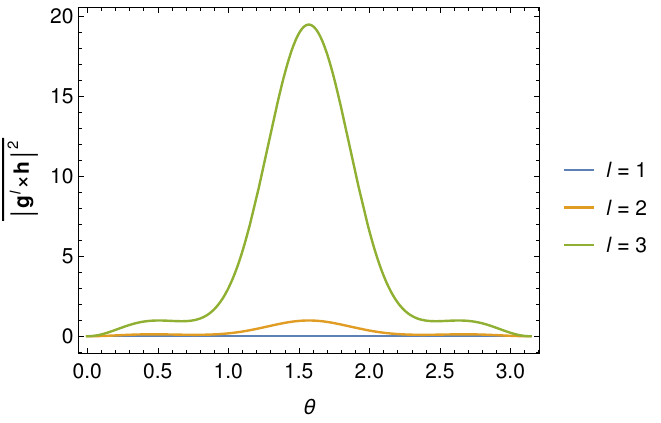}
\caption{\label{fig.glh}The $\phi$ averages $\overline{|\mathbf{g}^l \times \mathbf{h}|^2}$ characterizing the strength of the distortion terms $d^l_{E_u}$ that stabilize a distortion in the \textit{z} direction, as a function of the angle $\theta$. For details, see the text.}
\end{figure}

To compare the other two contributions, we note that the pairing state has rotational symmetry about the \textit{z} axis in the spherical approximation. This suggests to average $|\mathbf{g}^l \times \mathbf{h}|^2$ over the angle $\phi$, which yields
\begin{align}
\overline{|\mathbf{g}^1 \times \mathbf{h}|^2} &= 0 , \\
\overline{|\mathbf{g}^2 \times \mathbf{h}|^2}
  &= \frac{9}{23360}\, k^2 \sin^2 \theta\, \big(660 \cos 2 \theta + 1428 \cos 4 \theta \nonumber \\
&\quad{} + 108 \cos 6 \theta + 117 \cos 8 \theta + 1783\big) , \\
\overline{|\mathbf{g}^3 \times \mathbf{h}|^2}
  &= \frac{3}{1024}\, k^2 \sin^2 \theta\, \big({-}1004 \cos 2 \theta + 2324 \cos 4 \theta \nonumber \\
&\quad{} - 276 \cos 6 \theta + 213 \cos 8 \theta + 2839\big) .
\end{align}
These averages are plotted in Fig.\ \ref{fig.glh}. It is obvious that the third contribution is much larger than the second. Hence, the third term $d^3_{E_u}$ leads to a much stronger ASOC in the direction normal to the pseudomagnetic field $\mathbf{h}(\mathbf{k})$ compared to the second term $d^2_{E_u}$, while the first term is zero anyway. Therefore, the third term is most effective in stabilizing a distortion in the \textit{z} direction than the second and the second more than the first. This is exactly the order found numerically; see Table~\ref{tab.CC}.

\section{Pair-density wave}
\label{sec.PDW}

To address the question whether the system could show an instability toward a spatially modulated superconducting state---a PDW---we consider a Ginzburg-Landau description. We denote the complex superconducting amplitudes for the various pairing channels by $\Psi_m(\mathbf{r})$ and ignore the coupling to the electromagnetic field for simplicity. For our $J=3/2$ model, $m=0$ corresponds to the $A_{1g}$ pairing state, $m=1,2,3$ to $T_{2g}$, and $m=4,5$ to $E_g$. A PDW will occur generically if a term of first order in gradients is allowed in the Ginzburg-Landau free-energy density; see Ref.\ \cite{SSY17} for a discussion.

We here restrict ourselves to the case of $T_{2g}$ pairing, where we have a three-component superconducting order parameter $(\Psi_1,\Psi_2,\Psi_3)$ and the $\Psi_m$ transform as irreducible tensor operators of the irrep $T_{2g}$ of the spatial point group $m\bar{3}m = O_h$. The first-order gradient terms read as
\begin{equation}
f_\mathrm{grad} = \sum_{m,n=1}^3 \mathbf{a}_{mn} \cdot \Psi_m^* i\nabla \Psi_n .
\label{PDW.fgrad.10}
\end{equation}
From the reality of $f_\mathrm{grad}$, we obtain, using integration by parts, that $\mathbf{a}_{mn}^* = \mathbf{a}_{nm}$. The first-order terms are further constrained by the magnetic point-group symmetries. Under \emph{spatial} symmetries, the momentum operator $\hat{\bpi}=-i\bnabla$ transforms like the momentum $\mathbf{k}$. Furthermore, $\Psi_1$, $\Psi_2$, and $\Psi_3$ transform like the basis functions $k_yk_z$, $k_zk_x$, and $k_xk_y$, respectively. The same holds for their complex conjugates. The time-reversal transformation is more subtle since it involves a transposition: time reversal maps $\hat{\bpi}$ to $-\hat{\bpi}^T$ and $\Psi_n$ to $\Psi_n^*$. This implies the mapping
\begin{equation}
\Psi_m^* \hat{\bpi} \Psi_n \to - \Psi_m \hat{\bpi}^T \Psi_n^* = - \Psi_n^* \hat{\bpi} \Psi_m ,
\end{equation}
where we have used that the transposed momentum operator acts to the left instead of to the right. This is corroborated by the observation that $\hat{\bpi}$ must transform like the vector potential $\mathbf{A}$ to preserve gauge invariance and that $\Psi_m^* \mathbf{A} \Psi_n \to - \Psi_n^* \mathbf{A} \Psi_m$ under time reversal.

\begin{table}[h]
\caption{\label{tab.PDW.4b}Association of quantities relevant for the formation of a PDW with irreps of $m\bar{3}m1'$.}
\begin{ruledtabular}
\begin{tabular}{cc}
$\Psi_1^* \otimes \Psi_1 + \Psi_2^* \otimes \Psi_2 + \Psi_3^* \otimes \Psi_3$ & $A_{1g+}$ \\
$\Psi_1^* \otimes \Psi_1 - \Psi_2^* \otimes \Psi_2$ & \\
$\frac{1}{\sqrt{3}} \left( 2 \Psi_3^* \otimes \Psi_3 - \Psi_1^* \otimes \Psi_1 - \Psi_2^* \otimes \Psi_2 \right)$ &
  \raisebox{1.3ex}[-1.3ex]{$\Big\}\, E_{g+}$} \\
$\Psi_2^* \otimes \Psi_3 + \Psi_3^* \otimes \Psi_2$ & \\
$\Psi_3^* \otimes \Psi_1 + \Psi_1^* \otimes \Psi_3$ & \\
$\Psi_1^* \otimes \Psi_2 + \Psi_2^* \otimes \Psi_1$ & \raisebox{2.2ex}[-2.2ex]{$\bigg\}\, T_{2g+}$} \\
$\Psi_2^* \otimes \Psi_3 - \Psi_3^* \otimes \Psi_2$ & \\
$\Psi_3^* \otimes \Psi_1 - \Psi_1^* \otimes \Psi_3$ & \\
$\Psi_1^* \otimes \Psi_2 - \Psi_2^* \otimes \Psi_1$ & \raisebox{2.2ex}[-2.2ex]{$\bigg\}\, T_{1g-}$} \\ \hline
$\hat{\bpi} = -i\bnabla$ & $T_{1u-}$ \\ \hline
$\mathbf{p}$ & $T_{1u+}$ \\ \hline
$-\mathbf{p} \cdot i\bnabla$ & $A_{1g-}$ \\
$-p_x i\partial_x + p_y i\partial_y$ & \\
$-\frac{1}{\sqrt{3}} \left( 2 p_z i\partial_y - p_x i\partial_x - p_y i\partial_y \right)$ &
  \raisebox{1.3ex}[-1.3ex]{$\Big\}\, E_{g-}$} \\
$-p_y i\partial_z - p_z i\partial_y$ & \\
$-p_z i\partial_x - p_x i\partial_z$ & \\
$-p_x i\partial_y - p_y i\partial_x$ & \raisebox{2.2ex}[-2.2ex]{$\bigg\}\, T_{2g-}$} \\
$-p_y i\partial_z + p_z i\partial_y$ & \\
$-p_z i\partial_x + p_x i\partial_z$ & \\
$-p_x i\partial_y + p_y i\partial_x$ & \raisebox{2.2ex}[-2.2ex]{$\bigg\}\, T_{1g-}$}
\end{tabular}
\end{ruledtabular}
\end{table}

We can now analyze the transformation properties of bilinear products $\Psi_m^* \otimes \Psi_n$ under $m\bar{3}m1'$, where the Kronecker product indicates that the factors are not immediately multiplied as scalars but rather the momentum operator is inserted before they are contracted. We have found that $\Psi_m^* \otimes \Psi_n \to \Psi_n^* \otimes \Psi_m$ under time reversal. The combinations transforming as irreducible tensor operators of the irreps of $m\bar{3}m1'$ are given in Table \ref{tab.PDW.4b}. Note the appearance of $T_{1g-}$; the behavior under time reversal is nonstandard because of the transposition and cannot directly be inferred from the reduction $T_{2g+} \otimes T_{2g+} = A_{1g+} \oplus E_{g+} \oplus T_{1g+} \oplus T_{2g+}$.

The symmetries of momentum operators and distortional order parameters are also included in Table \ref{tab.PDW.4b}. In the harmonic regime, the coefficients $\mathbf{a}_{mn}$ in Eq.\ (\ref{PDW.fgrad.10}) are linear in the distortions $\mathbf{p}$. The irreps of all independent bilinear forms constructed from $\mathbf{p}$ and $-i\nabla$ are also given in Table~\ref{tab.PDW.4b}.

The free-energy density $f_\mathrm{grad}$ has to be invariant under $m\bar{3}m1'$. We can thus construct the allowed terms as all combinations of $\mathbf{p}$, $-i\bnabla$, and order-parameter bilinear forms that transform according to $A_{1g+}$. Since $\mathbf{p}$ is even under time reverse and $-i\bnabla$ is odd, the order-parameter bilinear has to be odd, which only permits the three combinations belonging to $T_{1g-}$. Hence, we find three allowed terms:
\begin{align}
f_{\mathrm{grad},1} &\propto - p_y \left( \Psi_2^* i\partial_z \Psi_3 - \Psi_3^* i\partial_z \Psi_2 \right)
  \nonumber \\
&\quad{} + p_z \left( \Psi_2^* i\partial_y \Psi_3 - \Psi_3^* i\partial_y \Psi_2 \right) , \\
f_{\mathrm{grad},2} &\propto - p_z \left( \Psi_3^* i\partial_x \Psi_1 - \Psi_1^* i\partial_x \Psi_3 \right)
  \nonumber \\
&\quad{} + p_x \left( \Psi_3^* i\partial_z \Psi_1 - \Psi_1^* i\partial_z \Psi_3 \right) , \\
f_{\mathrm{grad},3} &\propto - p_x \left( \Psi_1^* i\partial_y \Psi_2 - \Psi_2^* i\partial_y \Psi_1 \right)
  \nonumber \\
&\quad{} + p_y \left( \Psi_1^* i\partial_x \Psi_2 - \Psi_2^* i\partial_x \Psi_1 \right) .
\end{align}
Note that these terms do not fully agree with the Lifshitz invariants given in Ref.\ \cite{SSY17} for this pairing state.

We now make the ansatz
\begin{equation}
\Psi_m(\mathbf{r}) = \Psi_m^0\, e^{i\mathbf{Q}\cdot\mathbf{r}} .
\end{equation}
For the allowed gradient terms, this leads to
\begin{align}
f_{\mathrm{grad},1} &\propto (p_y Q_z - p_z Q_y) \left( \Psi_2^{0*} \Psi_3^0 - \Psi_3^{0*} \Psi_2^0 \right) , \\
f_{\mathrm{grad},2} &\propto (p_z Q_x - p_x Q_z) \left( \Psi_3^{0*} \Psi_1^0 - \Psi_1^{0*} \Psi_3^0 \right) , \\
f_{\mathrm{grad},3} &\propto (p_x Q_y - p_y Q_x) \left( \Psi_1^{0*} \Psi_2^0 - \Psi_2^{0*} \Psi_1^0 \right) .
\end{align}
So far, our results hold for any $T_{2g}$ order parameter. Specifically for the $\Delta\,(1,i,0)$ state, we obtain
\begin{align}
f_{\mathrm{grad},1} &= f_{\mathrm{grad},2} = 0 , \\
f_{\mathrm{grad},3} &\propto i\, |\Delta|^2\, (p_x Q_y - p_y Q_x) .
\end{align}
Noting that the average pseudomagnetic field is oriented along the \textit{z} direction and that it is proportional to $|\Delta|^2$ in the weak-coupling limit, we can write the remaining term as
\begin{equation}
f_{\mathrm{grad},3} \propto \mathbf{h} \cdot (\mathbf{p} \times \mathbf{Q})
  = (\mathbf{h} \times \mathbf{p}) \cdot \mathbf{Q}
\label{PDW.ffin.13}
\end{equation}
in that limit. An analysis based on the magnetic point group $4/mm'm'$ of the superconducting state leads to essentially the same result and is not presented here.

The linear term in Eq.\ (\ref{PDW.ffin.13}) would lead to a PDW with wave vector $\mathbf{Q} \propto \mathbf{h} \times \mathbf{p}$, evidently normal to both the pseudomagnetic field and the distortion. We only obtain a PDW if the distortion has a component normal to the pseudomagnetic field. Above, it has been concluded from numerical results and the pseudospin picture that the preferred distortion is parallel to the pseudomagnetic field. In that case, there is no modulation.

If a linear term exists the wavelength of the PDW scales with $1/p$ and thus diverges as the distortional transition temperature is approached from below. The weak-coupling distortional transition is thus expected to be unaffected by the formation of the PDW. However, a direct first-order transition toward a PDW with broken inversion symmetry could happen if a distortion not parallel to $\mathbf{h}$ is stable at all.

\section{Summary and conclusions}
\label{sec.conclusions}

In this paper, we have shown that the coupling of the electrons to the lattice always leads to a distortional weak-coupling instability of an in\-ver\-sion-sym\-me\-tric multiband superconductor that breaks time-reversal symmetry and has BFSs. At this instability, inversion symmetry is broken spontaneously and the BFSs are gapped out, except possibly in high-symmetry directions. In addition, the quasiparticle bands are shifted and are no longer even in momentum. If the shifts are sufficiently large, the state with broken inversion symmetry can still have BFSs \cite{Vol89,TSA17,YuF18,Vol18,LBH20,LiH20}. At zero temperature, the new gaps always lead to a reduction of the internal energy due to the appearance of a Cooper logarithm $\Phi^2\ln \Phi$ in terms of the absolute value of the distortional order parameter. A Cooper logarithm is characteristic for superconducting instabilities and it is remarkable that it here appears for a distortional instability. Its origin in both cases is that the quasiparticle bands are gapped at the Fermi energy. For the distortional instability, this is guaranteed by the $\mathcal{C}P$ symmetry of the undistorted state.

For the case where several distortional instabilities compete, we have introduced a relatively simple measure for their relative strengths. This measure is obtained by dividing the square of the change of the gap with the distortion by the normal-state Fermi velocity, and averaging the result over the BFSs. It is maximized by the leading instability and it discriminates highly selectively between competing instabilities since the energy gain depends exponentially on the measure.

In our example of a caesium chloride structure with $T_{2g}$ pairing with the order parameter $\Delta\,(1,i,0)$, the superconducting state lowers the cubic symmetry of the crystal to the tetragonal magnetic point group $4/mm'm'$ and there are competing distortions within and perpendicular to the $(001)$ plane. The leading instability depends on microscopic details of the system but a general argument based on an effective single-band model favors distortions in the \textit{z} direction. Two other pairing states that break time-reversal symmetry are plausible for this model \cite{BWW16,BAM18}: The $T_{2g}$ pairing state with order parameter $\Delta\, (1,\omega,\omega^2)$, $\omega=e^{2\pi i/3}$, has the trigonal magnetic point group $\bar{3}m'$ \cite{Lit13}. There are competing distortions within and perpendicular to the $(111)$ plane. The most plausible $E_g$ pairing state with order parameter $\Delta\, (1,i)$ has the magnetic point group $m\bar{3}m'$ \cite{Lit13}, which is still cubic. In this case, the distortions are isotropic in the harmonic approximation.

Actually, we have only demonstrated the instability for distortions that break inversion symmetry. We have pointed out that some structures with a basis do not permit uniform distortions that break inversion symmetry, the diamond structure being an example. However, it is always possible to break inversion symmetry by choosing a sufficiently large supercell---even if we start from a Bravais lattice we can take a supercell containing at least three lattice sites and deform it in such a way that the corresponding three-atom basis is itself not inversion symmetric. For commensurate distortions, the arguments made above carry over to the superlattice. In this case, the breaking of translation symmetry leads to the backfolding of bands and the gapping of band crossings already in the normal state. However, these new avoided crossings are generically not at the Fermi energy and thus do not affect the instability of the BFSs. We conclude that BFSs are unstable for \emph{any} structure. Only the induced distortion will necessarily not be uniform.

The distortional weak-coupling instability is universal and occurs even in the extreme limit where the band shifts induced by the distortion are everywhere larger than the induced band gaps so that the BFS are nowhere gapped out. The universality of the instability should be contrasted to the Peierls instability in one dimension \cite{Pei55,Fro54,Kup55}, for which the logarithm and thus the weak-coupling instability only exist if the system is tuned to half filling. It is also distinct from weak-coupling instabilities toward density-wave states in higher-dimensional systems that rely on fine-tuning of an approximate particle-hole symmetry which gives perfect nesting of Fermi pockets~\cite{FeM66,Ric70,MAS04,HCW08,CEE08,BrT09}.

We have also analyzed the instability in the pseudospin (single-band) picture. The resulting Hamiltonian can be mapped onto a dual model, in which the BFSs play the role of Fermi surfaces of a normal metal and the relevant component of the distortion becomes a \emph{superconducting} order parameter. This duality makes clear why the Cooper logarithm and the weak-coupling instability are generic. Finally, we have addressed the possibility of a PDW in the distorted state, based on methods discussed in Ref.\ \cite{SSY17}. Depending on the orientation of the distortional order parameter, a PDW can indeed form. However, its wavelength diverges at the distortional weak-coupling transition, rendering it irrelevant for the analysis of this transition.

It is important to note that other transitions can preempt the one discussed here. The most relevant possibility is that the superconducting state with broken time-reversal symmetry is replaced by a time-reversal-symmetric state at low temperatures because the growing BFSs and the associated density of states at the Fermi energy eventually become energetically disfavored \cite{MTB19}. In any case, the inversion-symmetric and time-re\-ver\-sal-sym\-me\-try-breaking state does not persist down to zero temperature.

\acknowledgements

The authors thank I. F. Herbut, J. M. Link, E.-G. Moon, and H. Oh for useful discussions. C.\,T. is grateful for financial support by the Deut\-sche For\-schungs\-ge\-mein\-schaft through the Collaborative Research Center SFB 1143, Project A04, the Research Training Group GRK 1621, and the Cluster of Excellence on Complexity and Topology in Quantum Matter ct.qmat (EXC~2147). P.\,M.\,R.\,B. was supported by the Marsden Fund Council from Government funding, managed by Royal Society Te Ap\={a}rangi.

\appendix

\section{Unitary symmetries of the BdG Hamiltonian}
\label{app.symmetry}

In this Appendix, we provide some background on unitary symmetries of the superconducting state. Assume that the normal state has a symmetry of the form $U H_N(R^{-1}\mathbf{k})\, U^\dagger = H_N(\mathbf{k})$, where $U$ is a unitary matrix acting on the Hilbert space of local degrees of freedom (here the effective spin) and $R$ is an $\mathrm{O}(3)$ matrix acting on momentum space. Evidently, $U$ is only fixed up to a phase factor. Extending the symmetry condition to Nambu space, we obtain
\begin{equation}
\mathcal{U}\, \mathcal{H}(R^{-1}\mathbf{k})\, \mathcal{U}^\dagger
  = \mathcal{H}(\mathbf{k}) ,
\label{eq.UHBdGU.3}
\end{equation}
with
\begin{equation}
\mathcal{U} = \begin{pmatrix}
    e^{i\alpha} U & 0 \\
    0 & e^{-i\alpha} U^*
  \end{pmatrix} ,
\end{equation}
where we have made the phase factor explicit. In the normal state ($\Delta=0$), this is equivalent to the original symmetry relation for any $\alpha$. In the superconducting state ($\Delta\neq 0$), the superconducting blocks in $\mathcal{H}(\mathbf{k})$ pick up factors of $e^{\pm 2i\alpha}$ under the transformation. Hence, Eq.\ (\ref{eq.UHBdGU.3}) cannot hold for all choices of $\alpha$. Rather, only two choices that differ by an overall sign of $U$ and thus of $\mathcal{U}$ are possible. This reflects the fact that superconductivity breaks a global $\mathrm{U}(1)$ invariance down to $\mathbb{Z}_2$~\cite{endnote.U1.breaking}. Typical $U$ in our case are rotation operators for the effective spin. It turns out that the allowed choices for $e^{i\alpha}$ are not necessarily the trivial ones $e^{i\alpha}=\pm 1$ and that they depend on the phase of $\Delta$.

\section{Symmetry constraints on band tilting and gapping of band crossings}
\label{app.constraints}

In this Appendix, we give a detailed discussion of the symmetries that can prevent asymmetric band shifts (tilting) or opening of gaps at band crossings.

(1) For certain (orientations of the) distortional order parameters and certain momenta $\mathbf{k}$, symmetry can force $\mathcal{H}_\mathrm{dist}(\mathbf{k})$ to vanish. For our model and distortions in the \textit{z} direction, note that for $\mathbf{k} = k_z\hat{\mathbf{z}}$ along the \textit{z} axis, i.e., for $k_x=k_y=0$, Table \ref{tab.kubasis} shows that only momentum basis functions belonging to $A_{1u}$ can be nonzero. But Eq.\ (\ref{eq.Hdist.pz}) does not contain any $A_{1u}$ basis functions. This implies that
\begin{equation}
H_{p_z}(k_z\hat{\mathbf{z}}) = 0 .
\end{equation}
This result can also be obtained as follows: $p_z$ transforms according to $A_{2u}$ and $k_z$ according to $A_{1u}$. Odd powers of $k_z$ also belong to $A_{1u}$, whereas even powers belong to $A_{1g}$. The relevant products are $A_{2u} \otimes A_{1u} = A_{2g}$ and $A_{2u} \otimes A_{1g} = A_{2u}$. There are no basis matrices belonging to either $A_{2g}$ or $A_{2u}$; see Table \ref{tab.basis}. Hence, no terms linear in $p_z$ and only depending on $k_z$ exist.

It is not true that $\mathcal{H}_\mathrm{dist}(\mathbf{k})$ always vanishes when $\mathbf{k}$ is parallel to $\mathbf{p}$: Consider the case that both $\mathbf{p}$ and $\mathbf{k}$ lie in the \textit{xy} plane and are oriented along the same principal axis, e.g., $\mathbf{p}=p_x\hat{\mathbf{x}}$ and $\mathbf{k}=k_x\hat{\mathbf{x}}$. Then Table \ref{tab.kubasis} shows that all odd one-dimensional momentum basis functions vanish, while for the $E_u$ basis functions only the first component vanishes, whereas the second is nonzero. Then the only remaining terms in Eq.\ (\ref{eq.Hdist.pxpy}) read as
\begin{align}
H_{p_x,0}(k_x\hat{\mathbf{x}}) &= \big[ 
  d_{E_u,2}^7(k_x\hat{\mathbf{x}})\, (J_xJ_yJ_z + J_zJ_yJ_x) \nonumber \\
&\quad{} + |\Delta|^2\, e_{E_u,2}^5(k_x\hat{\mathbf{x}})\, (J_xJ_y+J_yJ_x)
  \big] \, p_x .
\end{align}
These are the contributions from the $B_{2g}$ basis matrices. This makes sense since for $\mathbf{p}$ and $\mathbf{k}$ both parallel to $\hat{\mathbf{x}}$, the only nonvanishing combinations of the two belong to $A_{2g}$ and $B_{2g}$, as Table \ref{tab.EuEu.products} shows. But there are no basis matrices belonging to $A_{2g}$, see Table \ref{tab.basis}, and thus only $B_{2g}$ occurs. The result is that the electrons see the in-plane distortion even in this high-symmetry direction.

(2) A twofold rotation axis acts like the inversion symmetry in the plane normal to the axis. This pseudo-inversion can protect the symmetry of the energy spectrum and the existence of a $\mathbb{Z}_2$ invariant and thus of band crossings. Together, this implies that BFSs persist. In our example, the undistorted superconducting state possesses a twofold rotation symmetry $C_{2z}$ about the \textit{z} axis. The BdG Hamiltonian satisfies
\begin{equation}
\mathcal{U}_{C_{2z}}\, \mathcal{H}(-k_x,-k_y,k_z)\, \mathcal{U}_{C_{2z}}^\dagger
  = \mathcal{H}(\mathbf{k}) ,
\label{eq.C2zHBdG.3}
\end{equation}
with
\begin{align}
\mathcal{U}_{C_{2z}} &= \begin{pmatrix}
    -i\, e^{-i J_z \pi} & 0 \\
    0 & i\, e^{i J_z \pi}
  \end{pmatrix} \nonumber \\
&= \mathrm{diag}(1,-1,1,-1,1,-1,1,-1) ,
\label{eq.UC2z.3}
\end{align}
see also Appendix \ref{app.symmetry}. This transformation acts like inversion in the $k_z=0$ plane.

To see how the distortion affects this symmetry, we have to keep the distortional order parameter fixed under point-group transformations. $k_z$ is even under the rotation $C_{2z}$, while $(k_x,k_y)$ is odd. Then Table \ref{tab.kubasis} shows that momentum basis functions belonging to $A_{1u}$, $A_{2u}$, $B_{1u}$, and $B_{2u}$ are even, whereas functions belonging to $E_u$ are odd. The basis matrices in Table \ref{tab.basis} are even for the irreps $A_{1g}$, $B_{1g}$, and $B_{2g}$, whereas they are odd for the irrep $E_g$, as a look at a character table for $D_{4h}$ shows (note that $C_{2z}$ is not combined with time reversal in the magnetic point group). Equation (\ref{eq.Hdist.pz}) then shows that $H_{p_z}(\mathbf{k})$ is invariant under $C_{2z}$. This is not surprising since the distortion is parallel to the twofold axis. On the other hand, Eq.\ (\ref{eq.Hdist.pxpy}) shows that $H_{p_x,p_y}(\mathbf{k})$ is odd under the rotation $C_{2z}$, also as expected for a distortion normal to the twofold axis. Hence, for a distortion $\mathbf{p}=p_z\hat{\mathbf{z}}$ along the \textit{z} axis, the twofold rotation symmetry of the superconducting state survives. Since the rotation acts like inversion in the $k_z=0$ plane this implies that the spectrum remains symmetric in this plane for distortions $\mathbf{p}=p_z\hat{\mathbf{z}}$, i.e., there is no asymmetric shift of bands.

Charge-conjugation symmetry is described by $\mathcal{U}_C\, \mathcal{H}^T(-\mathbf{k})\, \mathcal{U}_C^\dagger = -\mathcal{H}(\mathbf{k})$, where $\mathcal{U}_C = \sigma_1 \otimes \boldone$ is the unitary part of the antiunitary charge conjugation operator $\mathcal{C}$. One easily checks that
\begin{equation}
(\mathcal{C}\mathcal{U}_{C_{2z}})^2 = \mathcal{U}_C \mathcal{U}_{C_{2z}}^* \mathcal{U}_C \mathcal{U}_{C_{2z}}
  = +\boldone .
\end{equation}
Hence, the effective inversion in the $k_z=0$ plane satisfies the condition for the existence of a $\mathbb{Z}_2$ invariant \cite{KST14,ZSW16,ABT17,BAM18}. Consequently, the band crossings survive in this plane and, since the crossing is not shifted away from zero energy, the BFS survives for $\mathbf{p}=p_z\hat{\mathbf{z}}$.

(3) Mirror symmetries act like the inversion symmetry on the axis normal to the mirror plane. This can protect the symmetry of the energy spectrum and the existence of a $\mathbb{Z}_2$ invariant and thus of band crossings. Together, this implies that BFSs persist. In our example, the undistorted superconducting state possesses a mirror symmetry $m_z$ with respect to the \textit{xy} plane. The BdG Hamiltonian satisfies
\begin{equation}
\mathcal{U}_{m_z}\, \mathcal{H}(k_x,k_y,-k_z)\, \mathcal{U}_{m_z}^\dagger
  = \mathcal{H}(\mathbf{k}) ,
\label{eq.mirrorHBdG.3}
\end{equation}
with
\begin{equation}
\mathcal{U}_{m_z} = \mathcal{U}_{C_{2z}} = \begin{pmatrix}
    -i\, e^{-i J_z \pi} & 0 \\
    0 & i\, e^{i J_z \pi}
  \end{pmatrix} .
\label{eq.Umz.3}
\end{equation}
This transformation acts like inversion on the $k_z$ axis, i.e., for $k_x=k_y=0$.

The remainder of the argument is very similar to the case of the twofold axis. To see how the distortion affects the mirror symmetry, we keep the distortional order parameter fixed. $k_z$ is odd under the mirror operation $m_z$, while $(k_x,k_y)$ is even. Then Table \ref{tab.kubasis} shows that momentum basis functions belonging to $A_{1u}$, $A_{2u}$, $B_{1u}$, and $B_{2u}$ are odd, whereas functions belonging to $E_u$ are even. The basis matrices in Table \ref{tab.basis} are even for the irreps $A_{1g}$, $B_{1g}$, and $B_{2g}$, whereas they are odd for the irrep $E_g$. Equation (\ref{eq.Hdist.pz}) then shows that $H_{p_z}(\mathbf{k})$ is odd under $m_z$, whereas Eq.\ (\ref{eq.Hdist.pxpy}) shows that $H_{p_x,p_y}(\mathbf{k})$ is even. Hence, for the distortion $\mathbf{p}$ lying in the \textit{xy} plane, the mirror symmetry of the superconducting state survives. This implies that the energy spectrum remains symmetric on the $k_z$ axis for in-plane distortions, i.e., there is no asymmetric shift of bands.

The combination with charge conjugation satisfies
\begin{equation}
(\mathcal{C}\mathcal{U}_{m_z})^2 = \mathcal{U}_C \mathcal{U}_{m_z}^* \mathcal{U}_C \mathcal{U}_{m_z} = +\boldone .
\end{equation}
Hence, the effective inversion on the $k_z$ axis satisfies the condition for the existence of a $\mathbb{Z}_2$ invariant and the BFS survives in this one-dimensional space for $\mathbf{p}=(p_x,p_y,0)$.

In fact, together with the first point we obtain a stronger statement. We have seen that for our model $H_{p_z}(k_z\hat{\mathbf{z}})=0$ holds. This implies that the value of $p_z$ is irrelevant on the $k_z$ axis. We thus find the same results, namely symmetric spectrum and survival of BFSs on the $k_z$ axis, for \emph{any} distortion $\mathbf{p}=(p_x,p_y,p_z)$. This is an example of a ``higher symmetry than expected.''

(4) Twofold rotation axes can protect crossings on the rotation axis (not in the plane perpendicular to it) for special pairing states. Specifically, if the bands belonging to one of the eigenvalues of the rotation operator do not show any superconductivity, no gaps can open for them. This case is not discussed further since for our example it does not give additional conditions. In general, the arguments would be similar to the following discussion for mirror symmetries.

(5) Mirror symmetries can protect crossings within the mirror plane (not on the axis perpendicular to it) for special pairing states. Specifically, if the bands belonging to one of the eigenvalues of the mirror operator do not show any superconductivity, no gaps can open for them. However, the spectrum generically becomes asymmetric. In our example, the symmetry in Eq.\ (\ref{eq.mirrorHBdG.3}) restricted to this plane becomes
\begin{equation}
\mathcal{U}_{m_z}\, \mathcal{H}(k_x,k_y,0)\, \mathcal{U}_{m_z}^\dagger
  = \mathcal{H}(k_x,k_y,0) .
\end{equation}
This is a local symmetry in momentum space. It is equivalent to $\big[\mathcal{U}_{m_z}, \mathcal{H}(k_x,k_y,0)\big] = 0$. Hence, one can block diagonalize the BdG Hamiltonian as
\begin{equation}
\mathcal{H}(k_x,k_y,0)
  \longrightarrow \begin{pmatrix}
    \mathcal{H}_+(k_x,k_y) & 0 \\
    0 & \mathcal{H}_-(k_x,k_y)
  \end{pmatrix} ,
\end{equation}
where $\mathcal{H}_\pm(k_x,k_y)$ acts on the sectors belonging to the eigenvalues $\pm 1$ of $\mathcal{U}_{m_z}$. Since $\mathcal{U}_{m_z}$ is already diagonal this is just a reordering of the rows and columns. As we have seen in Ref.\ \cite{BAM18}, for our special $T_{2g}$ pairing state, only $\mathcal{H}_-(k_x,k_y)$ contains superconductivity. Therefore, $\mathcal{H}_+(k_x,k_y)$ and thus half of the bands in the $k_z=0$ plane have no superconducting gaps.

As noted above, the $m_z$ mirror symmetry persists for a distortion in the \textit{xy} plane. Hence, it remains true that half of the bands in the $k_z=0$ plane do not show a superconducting gap. However, all bands become asymmetric.

(6) Magnetic symmetry operations involving twofold rotation axes can act as effective time reversal on these axes. We take $C_{2x}' = C_{2x} \mathcal{T}$ as an example. The undistorted superconducting state has this symmetry. It acts as
\begin{equation}
\mathcal{U}_{C_{2x}'} \mathcal{H}^T(-k_x,k_y,k_z)\,
  \mathcal{U}_{C_{2x}'}^\dagger = \mathcal{H}(\mathbf{k}) ,
\label{eq.C2xpHBdG.3}
\end{equation}
with
\begin{align}
\mathcal{U}_{C_{2x}'} &= \mathcal{U}_{C_{2x}} \mathcal{U}_T
  = \begin{pmatrix}
    e^{-i J_x \pi} & 0 \\
    0 & e^{i J_x \pi}
  \end{pmatrix}
  \begin{pmatrix}
    e^{i J_y \pi} & 0 \\
    0 & e^{i J_y \pi}
  \end{pmatrix} \nonumber \\
&= \mathrm{diag}(-i,i,-i,i,i,-i,i,-i) .
\label{eq.UC2xp.3}
\end{align}
On the $k_x$ axis, i.e., for $k_y=k_z=0$, the transformation behaves like time reversal.

For the distortion terms, note that on the $k_x$ axis only the second component of the $E_u$ basis functions is nonzero. It is odd under $C_{2x}'$. Of the basis matrices, the ones belonging to $A_{1g}$ and $B_{1g}$ are even, the ones for $B_{2g}$ are odd, and the two components of $E_g$ basis functions are odd and even, respectively. We thus find
\begin{align}
H_{p_z}(k_x\hat{\mathbf{x}}) &= \big[ 
  d_{E_u,2}^1(k_x\hat{\mathbf{x}})\, J_y
  + d_{E_u,2}^2(k_x\hat{\mathbf{x}})\, J_y^3 \nonumber \\
&{} + d_{E_u,2}^3(k_x\hat{\mathbf{x}})\, (-J_y(J_z^2-J_x^2) - (J_z^2-J_x^2)J_y)
  \nonumber \\
&{} + |\Delta|^2\, e_{E_u,2}^1(k_x\hat{\mathbf{x}})\, (J_yJ_z + J_zJ_y) \big]\, p_z ,
\end{align}
which is odd under $C_{2x}'$ for fixed $p_z$. Hence, a distortion along the \textit{z} direction destroys the symmetry, as expected.

It is useful to write down the terms for $p_x$ and $p_y$ distortions separately. We find
\begin{align}
H_{p_x}(k_x\hat{\mathbf{x}}) &= \big[
  d_{E_u,2}^7(k_x\hat{\mathbf{x}})\, (J_xJ_yJ_z + J_zJ_yJ_x)
  \nonumber \\
&\quad{} + |\Delta|^2\, e_{E_u,2}^5(k_x\hat{\mathbf{x}})\, (J_xJ_y+J_yJ_x)
  \big]\, p_x ,
\end{align}
which is even, and
\begin{align}
H_{p_y}(k_x\hat{\mathbf{x}}) &= \big[
  d_{E_u,2}^4(k_x\hat{\mathbf{x}})\, J_z
  + d_{E_u,2}^5(k_x\hat{\mathbf{x}})\, J_z^3
  \nonumber \\
&\quad{} - d_{E_u,2}^6(k_x\hat{\mathbf{x}})\, [J_z(J_x^2-J_y^2) + (J_x^2-J_y^2)J_z]
  \nonumber \\
&\quad{} + |\Delta|^2\, e_{E_u,2}^2(k_x\hat{\mathbf{x}})\, \boldone 
  \nonumber \\
&\quad{} + |\Delta|^2\, e_{E_u,2}^3(k_x\hat{\mathbf{x}})\, (2J_z^2-J_x^2-J_y^2)
  \nonumber \\
&\quad{} - |\Delta|^2\, e_{E_u,2}^4(k_x\hat{\mathbf{x}})\, (J_x^2-J_y^2)
  \big] \, p_y ,
\end{align}
which is odd. Hence, the distorted systems retains the effective time-reversal symmetry if the distortion is along $\hat{\mathbf{x}}$. This symmetry together with charge conjugation then implies that the energy spectrum remains symmetric on the $k_x$ axis, even though inversion symmetry is broken. On the other hand, this type of symmetry does not prevent the gapping out of band crossings.

(7) Magnetic mirror symmetries can act as effective time-reversal symmetry within the mirror planes. We take $m_x' = m_x \mathcal{T}$ as an example. This symmetry acts as
\begin{equation}
\mathcal{U}_{m_x'} \mathcal{H}^T(k_x,-k_y,-k_z)\,
  \mathcal{U}_{m_x'}^\dagger = \mathcal{H}(\mathbf{k}) ,
\label{eq.mxpHBdG.3}
\end{equation}
with $\mathcal{U}_{m_x'} = \mathcal{U}_{C_{2x}'}$, i.e., like time reversal in the $k_yk_z$ ($k_x=0$) plane.

In the distortion terms, note that in the $k_x=0$ plane only momentum basis functions belonging to $A_{1u}$, $B_{1u}$, and the first component of $E_u$ appear. They are odd under $m_x'$. The basis matrices belonging to $A_{1g}$ and $B_{1g}$ are even, the ones for $B_{2g}$ are odd, and the two components of $E_g$ basis functions are odd and even, respectively. We thus find that
\begin{widetext}
\begin{align}
H_{p_z}(0,k_y,k_z) &= \big[
  d_{B_{1u}}^1(0,k_y,k_z)\, (J_xJ_yJ_z + J_zJ_yJ_x)
  + d_{E_u,1}^1(0,k_y,k_z)\, J_x
  + d_{E_u,1}^2(0,k_y,k_z)\, J_x^3 \nonumber \\
& \quad{}+ d_{E_u,1}^3(0,k_y,k_z)\, [J_x(J_y^2-J_z^2) + (J_y^2-J_z^2)J_x]
  \nonumber \\
& \quad{}+ |\Delta|^2\, e_{B_{1u}}^1(0,k_y,k_z)\, (J_xJ_y+J_yJ_x)
  + |\Delta|^2\, e_{E_u,1}^1(0,k_y,k_z)\, (J_zJ_x + J_xJ_z) \big]\, p_z
\end{align}
is even under $m_x'$. Similarly,
\begin{align}
H_{p_x}(0,k_y,k_z) &= \big[ d_{E_u,1}^4(0,k_y,k_z)\, J_z
  + d_{E_u,1}^5(0,k_y,k_z)\, J_z^3
  \nonumber \\
& \quad{}+ d_{E_u,1}^6(0,k_y,k_z)\, [J_z(J_x^2-J_y^2) + (J_x^2-J_y^2)J_z]
  \nonumber \\
& \quad{}+ d_{A_{1u}}^1(0,k_y,k_z)\, J_y
  + d_{A_{1u}}^2(0,k_y,k_z)\, J_y^3 \nonumber \\
& \quad{}- d_{A_{1u}}^3(0,k_y,k_z)\,
    [J_y(J_z^2-J_x^2)+(J_z^2-J_x^2)J_y]
  + d_{B_{1u}}^2(0,k_y,k_z)\, J_y \nonumber \\
& \quad{}+ d_{B_{1u}}^3(0,k_y,k_z)\, J_y^3
  + d_{B_{1u}}^4(0,k_y,k_z)\, [-J_y(J_z^2-J_x^2)-(J_z^2-J_x^2)J_y]
  \nonumber \\
& \quad{}+ |\Delta|^2\, e_{E_u,1}^2(0,k_y,k_z)\, \boldone 
  + |\Delta|^2\, e_{E_u,1}^3(0,k_y,k_z)\, (2J_z^2-J_x^2-J_y^2)
  \nonumber \\
& \quad{}+ |\Delta|^2\, e_{E_u,1}^4(0,k_y,k_z)\, (J_x^2-J_y^2)
  + |\Delta|^2\, e_{A_{1u}}^1(0,k_y,k_z)\, (J_yJ_z+J_zJ_y)
  \nonumber \\
& \quad{}
  + |\Delta|^2\, e_{B_{1u}}^2(0,k_y,k_z)\, (J_yJ_z+J_zJ_y)
  \big] \, p_x
\end{align}
is odd and
\begin{align}
H_{p_y}(0,k_y,k_z) &= \big[
  d_{E_u,1}^7(0,k_y,k_z)\, (J_xJ_yJ_z + J_zJ_yJ_x)
  - d_{A_{1u}}^1(0,k_y,k_z)\, J_x
  - d_{A_{1u}}^2(0,k_y,k_z)\, J_x^3 \nonumber \\
& \quad{}- d_{A_{1u}}^3(0,k_y,k_z)\, [J_x(J_y^2-J_z^2)+(J_y^2-J_z^2)J_x]
  + d_{B_{1u}}^2(0,k_y,k_z)\, J_x \nonumber \\
& \quad{}+ d_{B_{1u}}^3(0,k_y,k_z)\, J_x^3
  + d_{B_{1u}}^4(0,k_y,k_z)\, [J_x(J_y^2-J_z^2)+(J_y^2-J_z^2)J_x]
  \nonumber \\
& \quad{}+ |\Delta|^2\, e_{E_u,1}^5(0,k_y,k_z)\, (J_xJ_y+J_yJ_x)
  - |\Delta|^2\, e_{A_{1u}}^1(0,k_y,k_z)\, (J_zJ_x+J_xJ_z)   \nonumber \\
& \quad{}+ |\Delta|^2\, e_{B_{1u}}^2(0,k_y,k_z)\, (J_zJ_x+J_xJ_z)
  \big]\, p_y
\end{align}
\end{widetext}
is even. Hence, a distortion in the \textit{yz} magnetic mirror plane respects the effective time-reversal symmetry for $\mathbf{k}$ in the same plane. Importantly, $\mathbf{p}$ and $\mathbf{k}$ need not be related beyond lying in the same plane. Together with charge conjugation this implies that the energy spectrum remains symmetric in this plane. Band crossings are not protected.

\section{Electronic internal energy}
\label{app.instability}

In this Appendix, we evaluate the change in the electronic internal energy due to distortions. We first consider the case that there are odd-in-mo\-men\-tum band shifts but no splittings. For the undistorted superconductor, inversion symmetry implies that the spectrum is even in momentum and it is possible to enumerate the eigenvalues in such a way that $E_n(-\mathbf{k}) = E_n(\mathbf{k})$ for $n=1,\ldots,2n_b$. Equation (\ref{Clog.Enext.3}) then implies that $E_{n_b+n}(\mathbf{k}) = -E_n(\mathbf{k})$ for $n\le n_b$ and $E_{n-n_b}(\mathbf{k}) = -E_n(\mathbf{k})$ for $n>n_b$. Thus the spectrum is symmetric at each $\mathbf{k}$ and the absolute values $|E_n(\mathbf{k})|$ are twofold degenerate.

Adding an odd correction $\delta E_n(\mathbf{k})$, we obtain new eigenenergies
\begin{equation}
\tilde E_n(\mathbf{k}) = E_n(\mathbf{k}) + \delta E_n(\mathbf{k}) .
\end{equation}
We have $\tilde E_n(-\mathbf{k}) = E_n(\mathbf{k}) - \delta E_n(\mathbf{k})$ and charge-con\-ju\-ga\-tion symmetry implies that there is another eigenenergy
\begin{equation}
\tilde E_{n'}(\mathbf{k}) = -\tilde E_n(-\mathbf{k})
  = -E_n(\mathbf{k}) + \delta E_n(\mathbf{k})
\end{equation}
at $\mathbf{k}$, where $n' = n_b+n$ for $n\le n_b$ and $n' = n-n_b$ for $n>n_b$. Importantly, the unperturbed energy of band $n'$ has opposite sign compared to the energy of band $n$ but their shift is the same. Let $E_n(\mathbf{k})\ge 0$ and $E_{n'}(\mathbf{k})\le 0$ without loss of generality. Then there are two cases: In region A, defined by $|\delta E_n(\mathbf{k})| \le E_n(\mathbf{k})$, the absolute values are
\begin{align}
|\tilde E_n(\mathbf{k})| &= E_n(\mathbf{k}) + \delta E_n(\mathbf{k}) , \\
|\tilde E_{n'}(\mathbf{k})| &= E_n(\mathbf{k}) - \delta E_n(\mathbf{k}) .
\end{align}
The shift of the absolute values appearing in $U_\mathrm{el}$ is then opposite for the two bands and drops out when the sum over bands is performed.

In region B with $|\delta E_n(\mathbf{k})| > E_n(\mathbf{k})$, the absolute values depend on the sign of $\delta E_n(\mathbf{k})$. We find
\begin{align}
|\tilde E_n(\mathbf{k})| &= \sgn(\delta E_n(\mathbf{k}))\, E_n(\mathbf{k})
  + |\delta E_n(\mathbf{k})| , \\
|\tilde E_{n'}(\mathbf{k})| &= -\sgn(\delta E_n(\mathbf{k}))\, E_n(\mathbf{k})
  + |\delta E_n(\mathbf{k})| .
\end{align}
The unperturbed values thus cancel in $U_\mathrm{el}$ but the shifts do not.

The shift of the electronic internal energy reads as
\begin{align}
\delta U_\mathrm{el} &= -\frac{1}{4} \! \sum_{(\mathbf{k},n)\,\in\,\mathrm{B}}
  | E_n(\mathbf{k}) + \delta E_n(\mathbf{k}) |
  + \frac{1}{4} \! \sum_{(\mathbf{k},n)\,\in\,\mathrm{B}}
  | E_n(\mathbf{k}) | \nonumber \\
&= \frac{1}{4} \! \sum_{(\mathbf{k},n)\,\in\,\mathrm{B}} \big[
  {-}|\delta E_n(\mathbf{k})| + |E_n(\mathbf{k})| \big] < 0 ,
\end{align}
where the sum is only over the region B for each band $n=1,\ldots,2n_b$ since in region A the shifts drop out. $\delta U_\mathrm{el}$ is negative since $|\delta E_n(\mathbf{k})| > |E_n(\mathbf{k})|$ in region B.

For small distortions, the change in the electronic internal energy is dominated by states close to the Fermi energy, i.e., close to the BFSs of the undistorted superconductor. In particular, region B is restricted to the vicinity of the BFSs. We can therefore expand the unperturbed quasiparticle energies and the odd corrections about momenta $\mathbf{k}_F$ on the BFSs:
\begin{align}
E_n(\mathbf{k}) &\cong \mathbf{v}_n(\mathbf{k}_F)
  \cdot (\mathbf{k}-\mathbf{k}_F) , \\
\delta E_n(\mathbf{k}) &\cong \delta E_n(\mathbf{k}_F) .
\end{align}
Here, $\mathbf{v}_n(\mathbf{k}_F) \equiv \partial E_n(\mathbf{k})/\partial\mathbf{k}|_{\mathbf{k}=\mathbf{k}_F}$ is the Fermi velocity of the quasiparticles at $\mathbf{k} = \mathbf{k}_F$. We now go over to a momentum integral and write
\begin{align}
\delta U_\mathrm{el}
&\cong \frac{V}{4}\! \sum_{n=1}^{2n_b} \int_\mathrm{B} \frac{d^3k}{(2\pi)^3} \big[
  | \mathbf{v}_n(\mathbf{k}_F) \cdot (\mathbf{k}-\mathbf{k}_F) |
  - |\delta E_n(\mathbf{k}_F)| \big] \nonumber \\
&= \frac{V}{32\pi^3} \sum_{n=1}^{2n_b} \int_\mathrm{BFS} d^2k_F
  \int_{|v_n(\mathbf{k}_F)\,k_\perp|\: \le\:
    |\delta E_n(\mathbf{k}_F)|} \!\!\! dk_\perp \nonumber \\
&\quad{} \times \big[ | v_n(\mathbf{k}_F)\, k_\perp |
  - |\delta E_n(\mathbf{k}_F)| \big] ,
\label{Clog.Uel.8}
\end{align}
where $v_n(\mathbf{k}_F) = |\mathbf{v}_n(\mathbf{k}_F)|$, $k_\perp = |\mathbf{k}_\perp|$, and $\mathbf{k}_\perp = \mathbf{k}-\mathbf{k}_F$, chosen normal to the BFS, i.e., parallel to $\mathbf{v}_n(\mathbf{k}_F)$. The integral $\int_\mathrm{BFS}$ is over the full BFS. The Jacobian equals unity since the transformation is a pure rotation. Evaluating $\delta U_\mathrm{el}$ further, we obtain
\begin{align}
\delta U_\mathrm{el} &\cong
  \frac{V}{32\pi^3} \sum_{n=1}^{2n_b} \int_\mathrm{BFS} d^2k_F\, \bigg[ v_n(\mathbf{k}_F) k_\perp^2
    \Big|_0^{|\delta E_n(\mathbf{k}_F)|/v_n(\mathbf{k}_F)} \nonumber \\
&\quad{} - 2 |\delta E_n(\mathbf{k}_F)|\,
    \frac{|\delta E_n(\mathbf{k}_F)|}{v_n(\mathbf{k}_F)} \bigg] \nonumber \\
&= -\frac{V}{32\pi^3} \sum_{n=1}^{2n_b} \int_\mathrm{BFS} d^2k_F\,
  \frac{\delta E_n^2(\mathbf{k}_F)}{v_n(\mathbf{k}_F)} .
\end{align}
This contribution is finite, negative, and of second order in the magnitude $\Phi=|\bPhi|$ of the distortions. This scaling results from one power of $\Phi$ from the size of the region B and another power from the energy reduction within this region. The result shows that the tilting of the bands alone would lead to a strong-coupling instability: The prefactor of $\Phi^2$ must be sufficiently large in magnitude to overcome the elastic energy.

Next, we treat the case that the distortion gaps out the BFSs but leaves the spectrum symmetric. Then for any perturbed band energy $\tilde E_n(\mathbf{k})$, there is an $n'$ with $\tilde E_{n'}(\mathbf{k}) = -\tilde E_n(\mathbf{k})$. We take $\tilde E_n(\mathbf{k})\ge 0$ and $\tilde E_{n'}(\mathbf{k})\le 0$, without loss of generality. In the vicinity of the unperturbed BFSs of bands $n$ and $n'$, we then have
\begin{align}
\tilde E_n(\mathbf{k}) &\cong \sqrt{v_n^2(\mathbf{k}_F)\, k_\perp^2
  + \eta_n^2(\mathbf{k}_F)} , \\
\tilde E_{n'}(\mathbf{k}) &\cong -\sqrt{v_n^2(\mathbf{k}_F)\, k_\perp^2
  + \eta_n^2(\mathbf{k}_F)} ,
\end{align}
where $\eta_n(\mathbf{k}_F)>0$ is half the \emph{distortional} energy gap at momentum $\mathbf{k}_F$ on the unperturbed BFS. We again split the integration in $U_\mathrm{el}$ into an integral over the BFSs and an integral over $k_\perp$ in the direction perpendicular to it. However, now the integration is not restricted to some part of the Brillouin zone. Moreover, the $k_\perp$ integral diverges if extended to $\mathbb{R}$. It thus requires a cutoff $\Lambda$, which, in principle, comes from high-energy physics not included in the expansion and also depends on $\mathbf{k}_F$. The situation is very similar to the superconducting instability and like in standard BCS theory we treat $\Lambda$ as a constant, which does not qualitatively affect the results, and write
\begin{align}
\delta U_\mathrm{el} &\cong \frac{V}{32\pi^3} \sum_{n=1}^{2n_b} \int_\mathrm{BFS}
  d^2k_F \int_{-\Lambda}^\Lambda dk_\perp \nonumber \\
&\quad{} \times \Big[
  {-}\sqrt{v_n^2(\mathbf{k}_F)\, k_\perp^2 + \eta_n^2(\mathbf{k}_F)}
  + v_n(\mathbf{k}_F)\, |k_\perp| \Big] \nonumber \\
&= \frac{V}{32\pi^3} \sum_{n=1}^{2n_b} \int_\mathrm{BFS}
  d^2k_F\ v_n(\mathbf{k}_F) \int_{-\Lambda}^\Lambda dk_\perp \nonumber \\
&\quad{} \times \left[ |k_\perp| - \sqrt{k_\perp^2
    + \frac{\eta_n^2(\mathbf{k}_F)}{v_n^2(\mathbf{k}_F)}} \right] .
\end{align}
Evaluation of the $k_\perp$ integral gives
\begin{align}
\delta U_\mathrm{el} &\cong \frac{V}{32\pi^3} \sum_{n=1}^{2n_b} \int_\mathrm{BFS}
  d^2k_F\ v_n(\mathbf{k}_F) \left[ \rule{0ex}{4.5ex} \Lambda\, \Bigg(\Lambda \right. \nonumber \\
&\quad \left. {} - \sqrt{\Lambda^2 + \frac{\eta_n^2(\mathbf{k}_F)}{v_n^2(\mathbf{k}_F)}}\Bigg)
  + \frac{\eta_n^2(\mathbf{k}_F)}
    {v_n^2(\mathbf{k}_F)}\, \ln
    \frac{\frac{\eta_n(\mathbf{k}_F)}{v_n(\mathbf{k}_F)}}
    {\Lambda + \frac{\eta_n^2(\mathbf{k}_F)}{v_n^2(\mathbf{k}_F)}} \right] \nonumber \\
&\cong \frac{V}{32\pi^3} \sum_{n=1}^{2n_b} \int_\mathrm{BFS}
  d^2k_F\, \bigg[ \left( {-} \ln 2 - \frac{1}{2} \right)
    \frac{\eta_n^2(\mathbf{k}_F)}{v_n(\mathbf{k}_F)} \nonumber \\
&\quad{} + \frac{\eta_n^2(\mathbf{k}_F)}{v_n(\mathbf{k}_F)}\,
    \ln \frac{\eta_n(\mathbf{k}_F)}{v_n(\mathbf{k}_F) \Lambda} \bigg] ,
\end{align}
where we have used that $\Lambda$ is large. Since the gap $\eta_n(\mathbf{k}_F)$ is generically of first order in $\Phi$ the first term in the brackets is again of order $\mathcal{O}(\Phi^2)$. However, the second term has an additional logarithmic factor. This Cooper logarithm leads to a weak-coupling instability: $\delta U_\mathrm{el}$ can always be reduced by making $\bPhi$ and thus $\eta_n$ nonzero.

Rewriting $\delta U_\mathrm{el}$ as
\begin{align}
\delta U_\mathrm{el} &\cong \frac{V}{32\pi^3} \sum_{n=1}^{2n_b} \int_\mathrm{BFS}
  d^2k_F\, \frac{\eta_n^2(\mathbf{k}_F)}{v_n(\mathbf{k}_F)}\,
   \bigg[ {-} \ln 2 - \frac{1}{2} \nonumber \\
&\quad{} + \ln\frac{\eta_0}{v_n(\mathbf{k}_F) \Lambda}
    + \ln \frac{\eta_n(\mathbf{k}_F)}{\eta_0} \bigg] ,
\label{Clog.Uel.12}
\end{align}
where $\eta_0$ is an arbitrary constant reference energy, we see that the cutoff $\Lambda$ does not affect the Cooper logarithm but only renormalizes the coefficient of the $\mathcal{O}(\Phi^2)$ term. This term is expected to be dominated by the elastic energy.

Finally, we turn to the  generic case combining an odd-in-momentum band shift with the opening of gaps at the band crossings. For each band, momentum space is split into two regions, where region A is defined by $|\delta E_n(\mathbf{k})| \le \sqrt{v_n^2(\mathbf{k}_F)\, k_\perp^2 + \eta_n^2(\mathbf{k}_F)}$ and region B by $|\delta E_n(\mathbf{k})| > \sqrt{v_n^2(\mathbf{k}_F)\, k_\perp^2 + \eta_n^2(\mathbf{k}_F)}$.

If the gap is larger than or equal to the shift, $\eta_n(\mathbf{k}_F) \ge |\delta E_n(\mathbf{k}_F)|$, then only region A exists. It is still true that the shift cancels out in region A so that we only obtain the contribution from the gap. If this holds everywhere on the BFSs, then Eq.\ (\ref{Clog.Uel.12}) for the electronic internal energy remains valid.

In the general case, where regions of type B exist, we define
\begin{equation}
k_c(\mathbf{k}_F) \equiv \frac{\sqrt{\delta E_n^2(\mathbf{k}_F)
  - \eta_n^2(\mathbf{k}_F)}}{v_n(\mathbf{k}_F)}
\end{equation}
for $|\delta E_n(\mathbf{k}_F)| \ge \eta_n(\mathbf{k}_F)$. If $k_c(\mathbf{k}_F)$ exists a band crosses the Fermi energy at $k_\perp = \pm k_c(\mathbf{k}_F)$. This means that the distorted superconductor still has BFSs. In region B, i.e., for $|k_\perp| < k_c(\mathbf{k}_F)$, we get a cancellation of the even parts of the perturbed bands analogous to Eq.\ (\ref{Clog.Uel.8}). The change in the electronic internal energy reads as
\begin{widetext}
\begin{align}
\delta U_\mathrm{el} &\cong \frac{V}{32\pi^3} \sum_{n=1}^{2n_b} \int_\mathrm{BFS}
  d^2k_F\, 2\, \Bigg\{ \int_0^{k_c(\mathbf{k}_F)} dk_\perp\, \big[
  {-}|\delta E_n(\mathbf{k}_F)| + v_n(\mathbf{k}_F)\, k_\perp \big] \nonumber \\
&\quad{}+ \int_{k_c(\mathbf{k}_F)}^\Lambda dk_\perp \left[
  {-}\sqrt{v_n^2(\mathbf{k}_F)\, k_\perp^2 + \eta_n^2(\mathbf{k}_F)}
  + v_n(\mathbf{k}_F)\, k_\perp \right] \Bigg\} \nonumber \\
&= \frac{V}{16\pi^3} \sum_{n=1}^{2n_b} \int_\mathrm{BFS} d^2k_F\, \Bigg\{
  \int_0^\Lambda dk_\perp\, v_n(\mathbf{k}_F)\, k_\perp
  - \int_0^{k_c(\mathbf{k}_F)} dk_\perp\, |\delta E_n(\mathbf{k}_F)|
  - \int_{k_c(\mathbf{k}_F)}^\Lambda dk_\perp\,
    \sqrt{v_n^2(\mathbf{k}_F)\, k_\perp^2 + \eta_n^2(\mathbf{k}_F)} \Bigg\} \nonumber \\
&\cong \frac{V}{32\pi^3} \sum_{n=1}^{2n_b} \int_\mathrm{BFS} d^2k_F\, \Bigg\{
  {-} 2|\delta E_n(\mathbf{k}_F)|\, k_c(\mathbf{k}_F)
  - \frac{\eta_n^2(\mathbf{k}_F)}{2v_n(\mathbf{k})}
  + k_c(\mathbf{k}_F)\,
    \sqrt{v_n^2(\mathbf{k}_F)\, k_c^2(\mathbf{k}_F) + \eta_n^2(\mathbf{k}_F)}
  \nonumber \\
&\quad{}+ \frac{\eta_n^2(\mathbf{k}_F)}{v_n(\mathbf{k}_F)}\,
    \ln \frac{v_n(\mathbf{k}_F)\, k_c(\mathbf{k})
    + \sqrt{v_n^2(\mathbf{k}_F)\, k_c^2(\mathbf{k}_F) + \eta_n^2(\mathbf{k}_F)}}
    {2 v_n(\mathbf{k}_F) \Lambda} \Bigg\} ,
\end{align}
where we have used that $\Lambda$ is large. We define the band shift in units of the gap as
\begin{equation}
\alpha_n(\mathbf{k}_F) \equiv \frac{|\delta E_n(\mathbf{k}_F)|}
  {\eta_n(\mathbf{k}_F)}
\end{equation}
so that
\begin{equation}
k_c(\mathbf{k}_F) \equiv \left\{ \begin{array}{ll}
  \displaystyle \frac{\eta_n(\mathbf{k}_F)\,
    \sqrt{\alpha_n^2(\mathbf{k}_F) - 1}}{v_n(\mathbf{k}_F)} &
  \mbox{for $\alpha_n(\mathbf{k}_F) > 1$,} \\
  0 & \mbox{for $\alpha_n(\mathbf{k}_F) \le 1$.}
  \end{array} \right.
\end{equation}
We now split the integral over the unperturbed BFSs into regions which $\alpha_n(\mathbf{k}_F) \gtrless 1$ and obtain
\begin{align}
\delta U_\mathrm{el} &\cong \frac{V}{32\pi^3} \sum_{n=1}^{2n_b} \Bigg\{ \int_\mathrm{BFS}
  d^2k_F\, \left( - \ln 2 - \frac{1}{2} \right)
    \frac{\eta_n^2(\mathbf{k}_F)}{v_n(\mathbf{k}_F)}
  + \int_{\mathrm{BFS}\:|\:\alpha_n(\mathbf{k}_F)\, \le\, 1} d^2k_F\,
    \frac{\eta_n^2(\mathbf{k}_F)}{v_n(\mathbf{k}_F)}\,
    \ln \frac{\eta_n(\mathbf{k}_F)}{v_n(\mathbf{k}_F)\Lambda} \nonumber \\
&\quad{}+ \int_{\mathrm{BFS}\:|\:\alpha_n(\mathbf{k}_F)\, >\, 1} d^2k_F\, \Bigg[
    \frac{\eta_n(\mathbf{k}_F)\,|\delta E_n(\mathbf{k}_F)|}
      {v_n(\mathbf{k}_F)}\, \sqrt{\alpha_n^2(\mathbf{k}_F)-1}
  + \frac{\eta_n^2(\mathbf{k}_F)}{v_n(\mathbf{k}_F)}\,
      \ln \frac{[\sqrt{\alpha_n^2(\mathbf{k}_F)-1} + \alpha_n(\mathbf{k}_F)]\,
        \eta_n(\mathbf{k}_F)}{v_n(\mathbf{k}_F)\Lambda} \Bigg] \Bigg\}
  \nonumber \\
&= \frac{V}{32\pi^3} \sum_{n=1}^{2n_b} \int_\mathrm{BFS}
  d^2k_F\, \frac{\eta_n^2(\mathbf{k}_F)}{v_n(\mathbf{k}_F)}
  \left[ - \ln 2 - \frac{1}{2} + u(\alpha_n(\mathbf{k}_F))
  + \ln \frac{\eta_n(\mathbf{k}_F)}{v_n(\mathbf{k}_F)\Lambda} \right]
  \nonumber \\
&= \frac{V}{32\pi^3} \sum_{n=1}^{2n_b} \int_\mathrm{BFS}
  d^2k_F\, \frac{\eta_n^2(\mathbf{k}_F)}{v_n(\mathbf{k}_F)}
  \left[ - \ln 2 - \frac{1}{2}
  + \ln \frac{\eta_0}{v_n(\mathbf{k}_F)\Lambda}
  + u(\alpha_n(\mathbf{k}_F))
  + \ln \frac{\eta_n(\mathbf{k}_F)}{\eta_0} \right] ,
\label{Clog.Uel.15}
\end{align}
with
\begin{equation}
u(\alpha) \equiv \left\{ \begin{array}{ll}
  \displaystyle \alpha \sqrt{\alpha^2-1}
    + \ln\Big(\sqrt{\alpha^2-1}+\alpha\Big) &
  \mbox{for $\alpha>1$,} \\[0.5ex]
  0 & \mbox{for $\alpha\le 1$.}
  \end{array} \right.
\end{equation}
The terms with indices $n$ and $n' = n \pm n_b$ contribute equally to the final sum in Eq.\ (\ref{Clog.Uel.15}) so that we can restrict the sum to $n=1,\ldots,n_b$ and multiply by two. This leads to the final result given in Eq.~(\ref{Clog.Uel.16}).
\end{widetext}


\begin{thebibliography}{99}
\bibitem{endnote.polar}In insulating crystals with caesium-chloride structure, for example CsCl itself, the optical phonons at the zone center are split into two degenerate transverse modes and a separate longitudinal mode. This is due to long-range Coulomb interactions in polar crystals and does not affect our analysis since we are dealing with conductors.
\bibitem{endnote.normal.limit}It is illuminating to consider the normal-state limit. Enumerating the quasiparticle energies in such a way that $E_n(\mathbf{k}) \to \xi_n(\mathbf{k})$ in the normal state, we obtain
\begin{displaymath}
U_\mathrm{el} \to -\frac{1}{2} \sum_{\mathbf{k},n} \big|\xi_n(\mathbf{k})\big|
  + \frac{1}{2} \sum_{\mathbf{k},n} \xi_n(\mathbf{k})
  = \!\! \sum_{\mathbf{k},n\:|\:\xi_n(\mathbf{k}) < 0} \!\! \xi_n(\mathbf{k}) .
\end{displaymath}
As expected, only the single-particle energies of occupied states contribute.
\bibitem{endnote.decoupling.term}The omitted mean-field-decoupling term is of second order in the superconducting order parameter $\Delta$, which in turn obtains second-order corrections from the distortion. Hence, corrections to the decoupling term start at order $\mathcal{O}(\Phi^4)$ and can be ignored.
\bibitem{endnote.U1.breaking}Without anomalous terms in the second-quantized Hamiltonian, all fermionic annihilation operators can be multiplied by $e^{i\alpha}$ and all creation operators by $e^{-i\alpha}$ without changing the physics. However, anomalous terms corresponding to products of two annihilation (or creation) operators are only invariant under multiplication of all fermionic operators by $\pm 1$. Hence $\mathrm{U}(1)$ is reduced to $\mathbb{Z}_2$.
\end{thebibliography}
\end{document}